\documentclass[ reprint,superscriptaddress, amsmath,amssymb, aps, pra,onecolumn]{revtex4-2}

\usepackage{graphicx}
\usepackage[english]{babel}

\begin{document}

\title{Diffuse optics for glaciology}

\author{Markus Allgaier}
\email{oqt@uoregon.edu} 
\affiliation{Department of Physics and Oregon Center for Optical, Molecular, and Quantum Science, University of Oregon, Eugene, Oregon 97403, USA}
\author{Brian J. Smith}
\affiliation{Department of Physics and Oregon Center for Optical, Molecular, and Quantum Science, University of Oregon, Eugene, Oregon 97403, USA}

\begin{abstract}
Optical probing of glaciers has the potential for tremendous impact on environmental science. However, glacier ice is turbid, which prohibits the use of most established optical measurements for determining a glacier's interior structure. Here, we propose a method for determining the depth, scattering and absorption length based upon diffuse propagation of short optical pulses. Our model allows us to extract several characteristics of the glacier. Performing Monte Carlo simulations implementing Mie scattering and mixed boundary conditions, we show that the proposed approach should be feasible with current technology. The results suggest that optical properties and geometry of the glacier can be extracted from realistic measurements, which could be implemented with low cost and small footprint.
\end{abstract}

\maketitle

\section{Introduction}

The optical properties of glacial ice encode information on paleoclimate \cite{bay2001,bramall2013}, ice crystal orientation and flow \cite{Weikusat2017,rongen2020,hellmann2021}, and hold the key to precise modeling of radiative transfer \cite{kokhanovsky2004,warren2006,cooper2020}. Optical remote sensing techniques are commonly employed to monitor snow cover \cite{demuth2011}. While lidar cannot penetrate more than a few centimeters into snow \cite{demuth2011}, exact knowledge of optical properties of snow and ice is nevertheless required to achieve high precision \cite{cooper2020}. The transport of light in glacier ice is dominated by scattering from bubbles, grain boundaries and dust \cite{price1997}, while pure ice absorption is small in comparison. In deep glacier ice, diffuse optics have been employed to successfully measure scattering and absorption coefficients independently of each other from the shape of laser pulses transmitted through a large volume \cite{askebjer1995,ackermann2006,aartsen2013}. Understanding of diffuse optics is tantamount to perform such measurements. Penetrating, non-invasive optical measurements on the surface would simplify data collection and allow the usage of low-cost, portable off-the-shelf components. In addition, similar techniques are known in medical imaging as diffuse optical imaging \cite{durduran2010}, where appropriate boundary conditions for diffuse optics have long been known \cite{haskell1994}. The theory of diffuse transport of light with boundaries holds the key to measuring not only local optical properties, but also to extracting information about the volume geometry, such as the thickness of a glacier. Ice thickness is typically measured using radar-sounding\cite{plewes2001}. Such techniques are successfully used on cold glaciers, but only low-frequency systems, which are typically not commercially available, are of use for temperate glaciers \cite{plewes2001,rutishauser2016}. The use of commercially available optical components could provide similar performance at a much reduced system size and cost.

While many optical properties of snow and ice are well studied \cite{warren2006,kokhanovsky2004,libois2013,warren2019,beres2020,cooper2020}, precise data on the scattering and absorption properties of near-surface ice is scarce. Scattering coefficients of glacier ice has been measured in deep ice (\(>1000\)\,m) at the South Pole \cite{askebjer1995,ackermann2006}. By extrapolating data collected in the arctic region one can conclude that scattering near the surface is too strong for penetrating measurements \cite{price1997}. However, recent measurements in the Greenland ablation zone revealed that this is not the case in warmer regions, with scattering lengths on the order of \(1\)\,m \cite{cooper2020}. Recently extracted ice cores from the Rh\^one glaciers also show a low density of bubbles of small size \cite{hellmann2021}, corroborating the order of magnitude of scattering observed in Greenland. This prompts a re-evaluation of the feasibility of a penetrating optical measurement technique.


In this paper we present a simple model for diffusive transport of light in a shallow glacier and outline the very basic principles needed to describe diffuse optical measurements of glacier ice. We describe the glacier as a semi-infinite slab, where transport of light is governed by both absorption and scattering. We evaluate appropriate boundary conditions for both an idealized, totally reflective bottom termination and Fresnel-reflection at the top ice-air interface, which is known in the context of diffuse optical medical imaging \cite{haskell1994}. With these boundary conditions, we show that information about the geometry becomes easily accessible by constructing solutions to the diffusion equation using the well-known method of images \cite{bryan1890}. This serves as an extension of the infinite-volume random walk model developed for AMANDA \cite{askebjer1997}. In the second half of the paper we employ Monte Carlo simulations of a random walk to verify the validity of the model and give an estimate of measurement performance and limitations under realistic conditions. The results show that scattering and absorption length are accessible through both time-resolved measurements and simple intensity measurements on the surface, while information about thickness and geometry in general are accessible with time-resolved measurements only. We discuss challenges arising for a possible experimental implementation, such as laser sources, detectors, realistic glacier ice structure and bedrock reflectivity. While more research is needed to understand all these details as well as test the robustness of the presented model, these results do outline the potential of diffuse optical measurements for glaciology.

\section{Theory}

\subsection{Optical properties of glacier ice}

The optical properties of glacier ice are governed by both the properties of pure ice, as well as the effect of trapped air bubbles and impurities. For example, the absorption coefficient \(\beta\) of pure ice, which attenuates transmitted light after length \(d\) by the factor \(\mathrm{exp}(-\beta d)\), is small in the visible spectrum, measurements give values between \(10^{-1}\mathrm{m}^{-1}\) in a laboratory setting to \(10^{-3}\mathrm{m}^{-1}\) in the field \cite{warren2006}. While measuring the exact value is a challenge, it is evident that pure ice is optically very clear, and absorption at wavelength above 600nm seems unaffected by impurities \cite{warren2006,cooper2020}. Transport of light is instead strongly governed by scattering at bubbles, impurities, crystal boundaries, and various other sources, with bubbles giving the dominant contribution \cite{price1997}. If the transmission through snow or ice is affected by both scattering and absorption, transmitted intensity \(I(d)\) at depth $d$ can be described by the attenuation coefficient \(\alpha=\sigma + \beta\):

\begin{equation}
    \frac{I(d)}{I_0} = e^{-\alpha d} = e^{-(\sigma + \beta)d},
    \label{eq:att}
\end{equation}

\noindent where $I_0$ is the input light intensity and \(\sigma\) is the isotropic scattering coefficient, which is much larger in snow than it is in ice. After a distance of \(1/\alpha\) in the ice, the initial intensity will have dropped to \(1/e\). The size and character of \(\sigma\) varies strongly throughout the literature. Attenuation, scattering and absorption coefficients have recently been measured in the Greenland ablation zone \cite{cooper2020}. Without snow cover in the ablation zone of a glacier, the bare ice is accessible to optical measurements and appears to be mostly uniform at depths of more than a few centimeters. The scattering and absorption coefficients were found to be of order \(\sigma \approx 1\mathrm{m}^{-1}\) and \(\beta \approx 10^{-2}...10^{-1}\mathrm{m}^{-1}\), respectively, corresponding to scattering and absorption length (mean free path) of \(1m\) and \(10...100m\). These findings suggest that \textit{diffusive} transport of light can be quite efficient, especially in the spectral region around 400\,nm. This implies that a small portion of light injected at the surface reaches deep into the volume, and some signature of the volume depth may be encoded in light detected at the surface.

From the existing literature it is evident that absorption in the visible spectrum depends strongly on the content of light absorbing impurities, while the physical scattering coefficient is largely independent of wavelength.
Independent measurement of \(\sigma\) and \( \beta\) without the use of reference values, i.e., not relying on the attenuation coefficient from Eq. \ref{eq:att}, is challenging. To the best of the authors' knowledge this has only been achieved in the framework of the random-flight model employed at AMANDA \cite{askebjer1997,ackermann2006}, where the distribution of photon arrival times between a source and detector embedded deep in the ice was measured.

Exact knowledge of \(\sigma\) and \( \beta\) are tantamount to establish a useful and realistic model of the transport of light in a glacier. If these parameters can be extracted from measurement, such a model can in turn be used to obtain information about the geometry of a glacier, namely its thickness. These questions are similar to diffuse optical imaging in the context of life science \cite{durduran2010}.

\subsection{Random flights: From multi-scattering to diffusion}

\begin{figure}
    \centering
    \includegraphics[width=0.95\textwidth]{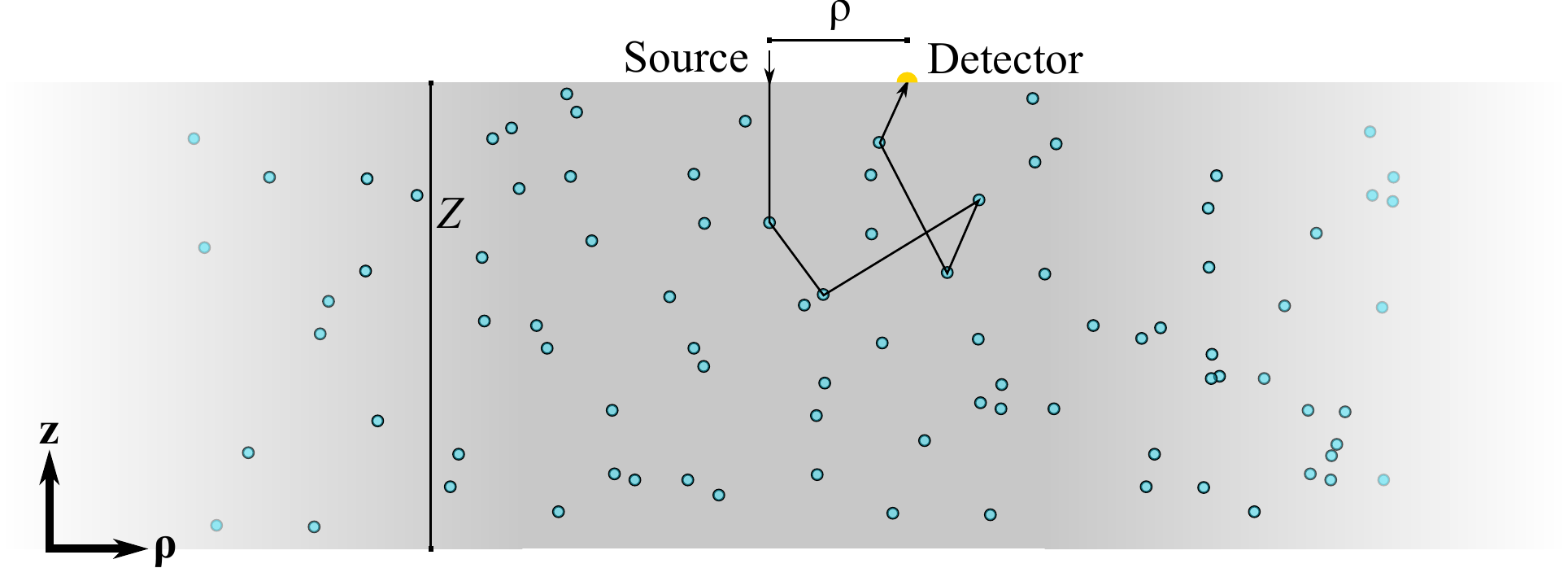}
    \caption{Sketch of transport of light in glacier ice of thickness \(Z\). Light gets injected downward from the surface and eventually gets scattered into a random direction by bubbles in the ice. Some light makes it back to the surface and gets detected by a detector separated from the source by the distance \(\rho\).}
    \label{fig:sketch}
\end{figure}

To derive a simple model of diffuse light transport in a glacier, we model the glacier as a semi-infinite slab.  The system is symmetric around the \(z\)-axis. The volume is uniformly turbid, meaning the scattering coefficient is larger than the absorption coefficient, and both, as well as the index of refraction \(n\), are assumed to be constant. The volume has a reflective boundary at the bottom (\(z=-Z\)), and a semi-reflective boundary at the top (\(z=0\)). An instantaneous (infinitely short pulsed) source of light is injected at \(z=\rho =0\), a detector is placed at the surface to collect light at the point of interest \(z=0,\rho\), where we are using cylindrical coordinates. We assume that light occupies a narrow spectral range and spectral dependencies can hence be neglected. The geometry is depicted in Figure \ref{fig:sketch}.

After injection, photons travelling on average for \(l_{\mathrm{sca}}=1/\sigma\), where they are scattered into a random direction, as prescribed by the probability density \(p(\theta)\) or scattering angle \(\theta\). The photon will travel a random distance \(r\) before it is scattered again. Each photon is scattered many times before reaching a boundary or being absorbed, performing a "random flight", while on average covering the distance \(l_{\mathrm{sca}}\) with every scattering step, but in a random direction. Finding the photon after a certain number of scattering steps becomes a statistical problem instead of an optical one that can be described simply by propagation.
The probability density \(P(R)\) to find a photon at distance \(R\) from the injection point after $N$ steps is described by the distribution \cite{chandrasekhar1943}

\begin{equation}
    P(R) = \frac{1}{\left(\frac{2}{3}\pi N\langle r^2 \rangle\right)^{3/2}}\mathrm{exp} \left( -3 R^2 / 2 N \langle r^2 \rangle \right),
\end{equation}

\noindent where \(\langle r^2 \rangle\) described the mean square displacement in the random flight. Here, we first establish the connection between the governing length scale of the random flight, \(\langle r^2 \rangle \), and the transport parameter of \textit{diffusion}, the diffusion constant \(D\). Using this parameter, we rewrite the above probability density into the well-known solution to the diffusion equation in an infinite volume. Again, it represents a probability density to find a photon at distance \(R\) from the source at time \(t\) after being emitted:

\begin{equation}
    P(R,t) = \frac{1}{\left(4\pi Dt\right)^{3/2}} \mathrm{exp} \left( -3  R^2 / 4Dt \right),
\end{equation}

\noindent with

\begin{equation}
    D = \frac{N \langle r^2 \rangle}{6 t} = \frac{c \langle r^2 \rangle}{6 l_{\mathrm{sca}}},
    \label{eq:diffconst}
\end{equation}

\noindent where we introduced the stepping rate \(N/t\). The distance traveled is \(N l_{\mathrm{sca}} = c t\) by design, with the speed of light in the medium, \(c\), which removes the apparent time-dependence of the diffusion constant. The diffusion constant is independent of the actual distribution of scattering lengths and depends only on a mean transport parameter \(\langle r^2 \rangle \).

Askebjer \textit{et al.}\cite{askebjer1997}  uses an additional dampening term to account for absorption. This term introduces a weight to components at long times. With the absorption term, we modify the previous density function. Since the probability to find the photon at large times is exponentially decreasing, the following density function is formally not a probability density function, because it does not integrate to 1, but it does still have the same form and information about where and when to find the photon. It reads

\begin{equation}
    P(R,t) = \frac{1}{\left(4\pi Dt\right)^{3/2}}\mathrm{exp} \left( -\frac{ R^2 }{ 4Dt } - \beta t\right).
    \label{eq:rw:infinite}
\end{equation}

\noindent This is a solution to the photon diffusion equation. Transitioning to \textit{macroscopic} quantities, from a probability density \(P(R,t)\) for the location of a single particle to a fluence rate \(\phi(R,t)\), describing the isotropic fluence rate in units of [${\rm{Wm}}^{-2}$], originating from an isotropic point source of unit energy. The diffusion-kernel with absorption is a well-known solution to the isotropic, infinite-volume case of the photon diffusion equation

\begin{equation}
    \frac{\partial \phi(R,t)}{\partial t} = D \nabla^2 \partial \phi(R,t) - c\beta\phi(R,t) + cS(R,t),
\end{equation}

\noindent where \(S(R,t)\) is the source term.

\subsection{The role of anisotropy}

Before we move on, it is worthwhile to inspect the diffusion constant \(D\) and the physical interpretation of the mean distance and mean square displacement. For a "truly random" flight, the scattering angle \(\theta\) is uniformly distributed. It is straightforward that mean distance and mean square displacement are then

\begin{equation}
    \langle r \rangle = \frac{1}{N}\sum_{i=1}^N r_i = 0,\quad \langle r^2 \rangle = \frac{1}{N}\left(\sum_{i=1}^N r_i \right)^2 = l_{\mathrm{sca}}^2.
    \label{eq:msd1}
\end{equation}

\noindent This implies that the distribution is always centered around the source point, since the average distance traveled is 0, while the mean square displacement, which is a measure of the extent of the distribution, is the scattering length squared. Scattering by bubbles in particular is not isotropic, but forward-peaked, with the average cosine of the scattering angle being \(\langle \cos \theta\rangle=0.75\) \cite{askebjer1997}. This is commonly known as the anisotropy factor first introduced in meteorology \cite{bohren1987} to describe attenuation in terms of an effective scattering constant

\begin{equation}
    \sigma_{\mathrm{eff}} = \sigma (1- \langle \cos \theta \rangle).
\end{equation}

\noindent This effective scattering coefficient can successfully describe attenuation in the sense of Eq. \ref{eq:att}, even when the distribution of scattering angles is not uniform, i.e., when scattering is anisotropic. While attenuation may behave the same for an effective scattering constant compared to an actual isotropic scattering constant, mean distance and mean square displacement differ\cite{askebjer1997}:

\begin{equation}
    \langle r \rangle = l_{\mathrm{sca}},\quad \langle r^2 \rangle = 2l_{\mathrm{sca}}^2
    \label{eq:msd2}
\end{equation}

\noindent This yields a factor of 2 difference for the diffusion constant between truly isotropic scattering and effective isotropic scattering. Using the mean square displacement for forward-peaked scattering and substituting into Eq. \ref{eq:diffconst} along with the effective scattering length we find that the diffusion constant for the case of scattering on bubbles or other forward-peaked scattering processes is

\begin{equation}
    D = \frac{c \langle r^2 \rangle}{6 l_{\mathrm{eff}}} = \frac{2c l_{\mathrm{eff}}^2}{6 l_{\mathrm{eff}}} = \frac{c l_{\mathrm{eff}}}{3}.
\end{equation}

\subsection{Boundary condition}

The volume we are employing to model the glacier has two boundaries. The top boundary is semi-reflective, the bottom boundary is reflective. As mentioned before, assuming unit reflectivity at the base is an oversimplification, but it allows for a straightforward, intuitive solution. This bottom boundary can easily be modeled using the method of images, which is well known in the context of electromagnetism and conduction of heat, which is also described by diffusion \cite{bryan1890}. The method is based on the assumption that any sum of solutions \(\phi_1,\phi_2\) to the diffusion equation is itself a solution to the diffusion equation

\begin{equation}
        \frac{\partial \left(\phi_1(R,t)+\phi_2(R,t)\right)}{\partial t} = D \nabla^2  \left(\phi_1(R,t)+\phi_2(R,t)\right) - c\beta \left(\phi_1(R,t)+\phi_2(R,t)\right) + cS(R,t).
\end{equation}

\noindent This follows from the linearity of the diffusion differential equation. The method of images allows the construction of a solution satisfying various boundary conditions from a source term, and several mirror sources. For a fully reflective barrier at \(z=-Z\), the mirror source term is identical to the source, but mirrored around the boundary, giving the solution with two source terms, where one is mirrored around the boundary and hence placed at \(z=-2Z\):

\begin{equation}
    \phi(R,t) = \frac{1}{\left(4\pi Dt\right)^{3/2}}\mathrm{exp} \left(-\frac{\rho^2}{4Dt}  - \beta t\right)\left(\mathrm{exp} \left(-\frac{z^2}{ 4Dt }\right) +  \mathrm{exp} \left(-\frac{(z+2Z)^2 }{4Dt}\right) \right).
\end{equation}

\noindent We simplify this expression by defining a point-source term as

\begin{equation}
\phi_{+}(z) = \frac{1}{\left(4\pi Dt\right)^{3/2}}\mathrm{exp} \left(-\frac{\rho^2}{4Dt}  - \beta t\right)\mathrm{exp} \left(-\frac{z^2}{ 4Dt }\right),
\end{equation}

\noindent which allows us to rewrite the above equation

\begin{equation}
    \phi(R,t) = \phi_{+}(z) + \phi_{+}(z+2Z).
\end{equation}

\noindent Here, it is assumed that the source is isotropic, which isn't the case for injection of a laser beam from the surface. We can approximate isotropic behavior by replacing the source (and mirror source) by effective sources, shifted by one effective scattering length into the slab:

\begin{equation}
    \phi(R,t) = \phi_{+}(z+l_{\mathrm{eff}}) +  \phi_{+}(z+2Z-l_{\mathrm{eff}}).
\end{equation}

This equation would be sufficient if all light escaped the top boundary, in which case it completely describes the time-dependent power density distribution in the slab in the presence of a reflective boundary. Unfortunately, behavior at the top boundary isn't that simple. To introduce a mixed boundary condition that describes the Fresnel reflections at the boundary between two media with different refractive index, ice and air, we follow the derivation by Haskell \textit{et al.}\cite{haskell1994}. To find a suitable expression, we need to introduce more radiometric quantities, namely the radiance \(L\) and flux \(\vec{j|}\). Radiance is a directional power density per unit area and therefore describes the directional transport of optical power. In a diffuse medium, the total radiance is composed of the isotropic fluence rate and a small directional flux:

\begin{equation}
    L(R,\vec{s},t) = \frac{\phi(R,t)}{4\pi} + \frac{3}{4\pi}\vec{j}(R,t)\cdot \vec{s}.
\end{equation}

\noindent The general diffusion approximation which allows us to describe the transport of light with such equations in the first place dictates that the radiance is dominated by the fluence rate, and the flux component is small. All outward radiation that is repelled by Fresnel reflection can be replaced by an inbound irradiance, analog to the method of images:

\begin{equation}
    L_{in} = R L_{out} = \iint_{\vec{s}\cdot\vec{n}>0} R(\vec{s})L(\vec{s})\vec{s}\cdot \vec{n} d\Omega,
\end{equation}

\noindent with

\begin{equation}
    R(\theta) = \frac{1}{2}\left(\frac{n \cos \theta' - \cos \theta}{n \cos \theta' + \cos \theta} \right)^2 +  \frac{1}{2}\left(\frac{n \cos \theta - \cos \theta'}{n \cos \theta + \cos \theta'} \right)^2, \quad \theta<\theta_c,
\end{equation}

\noindent where \(\theta_c\) is the critical angle above which total internal reflection occurs, \(\theta'\) is the external refracted angle. Note that we are averaging the Fresnel reflection for both linear polarizations, assuming that polarization is random after many scattering steps.

Left and right side of the reflected radiance equation evaluate as follows:

\begin{equation}
    \frac{\phi(R,t)}{4} + \frac{j_z}{2} = R_{\phi}\frac{\phi(R,t)}{4} - R_j\frac{j_z}{2},
\end{equation}

\noindent where the integrals for fluence rate and flux are

\begin{equation}
    R_{\phi} = \int_0^{\theta_c}2 \sin \theta \cos \theta R(\theta)d\theta,
\end{equation}
\begin{equation}
    R_{j} = \int_0^{\theta_c}3 \sin \theta \cos^2 \theta R(\theta)d\theta.
\end{equation}

\noindent With these definitions, we can write

\begin{equation}
    j_z = -\frac{1}{2}\frac{1-R_{\phi}}{1+R_j}\phi.
\end{equation}

\noindent Using Fick's law, which links fluence rate and flux in the diffusive regime, the boundary condition can finally be identified as

\begin{equation}
    -\frac{D}{c}\nabla \phi(R,t) = j = -\frac{1}{2}\frac{1-R_{\phi}}{1+R_j}\phi,
\end{equation}
\noindent with
\begin{equation}
    \phi(R,t) = \frac{2 l_{\mathrm{eff}}}{3}\frac{1+R_j}{1-R_{\phi}} \frac{\partial \phi}{\partial z} \equiv h\frac{\partial \phi}{\partial z}
\end{equation}

\noindent at the top of the volume (z=0). Ignoring the bottom boundary entirely for the moment, the mirror source's fluence rate \(\phi_m\) can be calculated using the boundary condition and infinite-volume solution \cite{bryan1890}:

\begin{align}
     \frac{\partial \phi_m}{\partial z} - \frac{\phi_m}{h} = -\left(\frac{\partial}{\partial z} - \frac{1}{h}\right)
    \frac{1}{\left(4\pi Dt\right)^{3/2}}\mathrm{exp} \left(-\frac{\rho^2}{4Dt}  - \beta t\right)\mathrm{exp} \left(-\frac{(z+l_{\mathrm{eff}})^2}{ 4Dt }\right) 
\end{align}

\noindent at \(z=0\), at which point we may use \(-(z+l_{\mathrm{eff}})=z-l_{\mathrm{eff}}\). Multiplying both sides by \(\mathrm{exp}(-z/h)\) gives:

\begin{align}
    \frac{\partial }{\partial z}\left( e^{-z/h}\phi_m \right)  = &\frac{\partial}{\partial z}\left( e^{-z/h}
    \frac{1}{\left(4\pi Dt\right)^{3/2}}\mathrm{exp} \left(-\frac{\rho^2}{4Dt}  - \beta t\right)\mathrm{exp} \left(-\frac{(z-l_{\mathrm{eff}})^2}{ 4Dt }\right)\right) \nonumber\\ 
    &\quad - \frac{2}{h} e^{-z/h} \frac{1}{\left(4\pi Dt\right)^{3/2}}\mathrm{exp} \left(-\frac{\rho^2}{4Dt}  - \beta t\right)\mathrm{exp} \left(-\frac{(z-l_{\mathrm{eff}})^2}{ 4Dt }\right).
\end{align}

This expression can be integrated, moving the new dampening terms \(\mathrm{exp}(-z/h)\) to the right side:

\begin{align}
   \phi_m (R,t) &=  \phi_+(z-l_{\mathrm{eff}}) - \frac{2}{h}\int_z^\infty e^{-(z-z')/h} \phi_+(z'-l_{\mathrm{eff}})dz' \nonumber\\
    & =   \phi_+(z-l_{\mathrm{eff}}) - \frac{2}{h}\int_0^\infty e^{-l/h} \phi_+(z-l-l_{\mathrm{eff}})dl,
    \label{eq:phi:infinite}
\end{align}

\noindent which consists of a mirror source at \(z=-l_{\mathrm{eff}}\), and an exponentially dampened line of sinks extending from \(z=-l_{\mathrm{eff}}\) to \(-\infty\).

So far, we have extended the infinite-volume description to non-isotropic injection and treated the two interfaces separately with different boundary conditions. For a general solution, both need to be combined. Since both interfaces are at least partially reflective, the solution will have to consist of an infinite sum of sources, sinks and their mirror images to account for the repeated reflections between both. We shall restrict the treatment of this problem to the surface, \(z=0\), and calculate the ratio between two consecutive round trips through the volume at the surface. Using a single source at \(z=0\), the ratio between the signal at \(\rho,z=0\) and the signal from a mirror source at \((\rho,z=-2Z)\) is then

\begin{equation}
    \frac{\phi_m}{\phi} = \mathrm{exp}\left( -\frac{Z^2}{Dt}  \right) = \mathrm{exp}\left( -\frac{3Z^2}{c l_{\mathrm{eff}}t}  \right).
\end{equation}

Assuming typical diffuse optical transit times \(ct\approx 1\)\,m and a sufficiently large volume to justify the diffuse treatment, \(Z/l_{\mathrm{eff}}\gg 1\), this ratio will be small. In addition, for the signal at the surface, only full integer reflections will contribute, and it is reasonable to only consider the first roundtrip reflection. Hence, we can write the full solution as the effective source distribution accounting for the partial boundary condition at the surface, plus a mirror distribution centered around \(z=-2Z\), consisting of 6 terms in total:

\begin{align}
   \phi (\rho,t,z=0) =   &\ \phi_+(l_{\mathrm{eff}}) + \phi_+(-l_{\mathrm{eff}})
   + \phi_+(l_{\mathrm{eff}} + 2Z) + \phi_+(-l_{\mathrm{eff}}+2Z)  \nonumber\\
     & - \frac{2}{h}\int_0^\infty e^{-l/h}\phi_{+}(-l-l_{\mathrm{eff}})dl  - \frac{2}{h}\int_0^\infty e^{-l/h}\phi_{+}(l+l_{\mathrm{eff}}+2Z) dl,
    \label{eq:theo:bounded}
\end{align}

\noindent once again using the compact definition of a source term \(\phi_+\). If higher precision is needed, the bottom mirror source distribution can be mirrored once more by applying the same method that was used to obtain the top source distribution, and so forth.

While constructing a mirror source image to account for a totally reflective boundary is both simple and instructive, it may not be applicable in many real-world applications. In the presence of loss at the lower boundary, introducing a mixed boundary may be more appropriate. If optical properties of the interface are known, this can be accomplished by following the derivation shown here but may prove challenging in the presence of a lossy medium. Employing a so-called extrapolated boundary \cite{haskell1994}, which involves shifting the mirror source image further away from the physical boundary, may prove valuable to approximate a lossy interface sufficiently. We discuss realistic values for the interface's reflectivity in section \ref{sec:exp}.

\subsection{Instantaneous source and time-resolved measurement}

\begin{figure}
    \centering
    \includegraphics[width=0.49\textwidth]{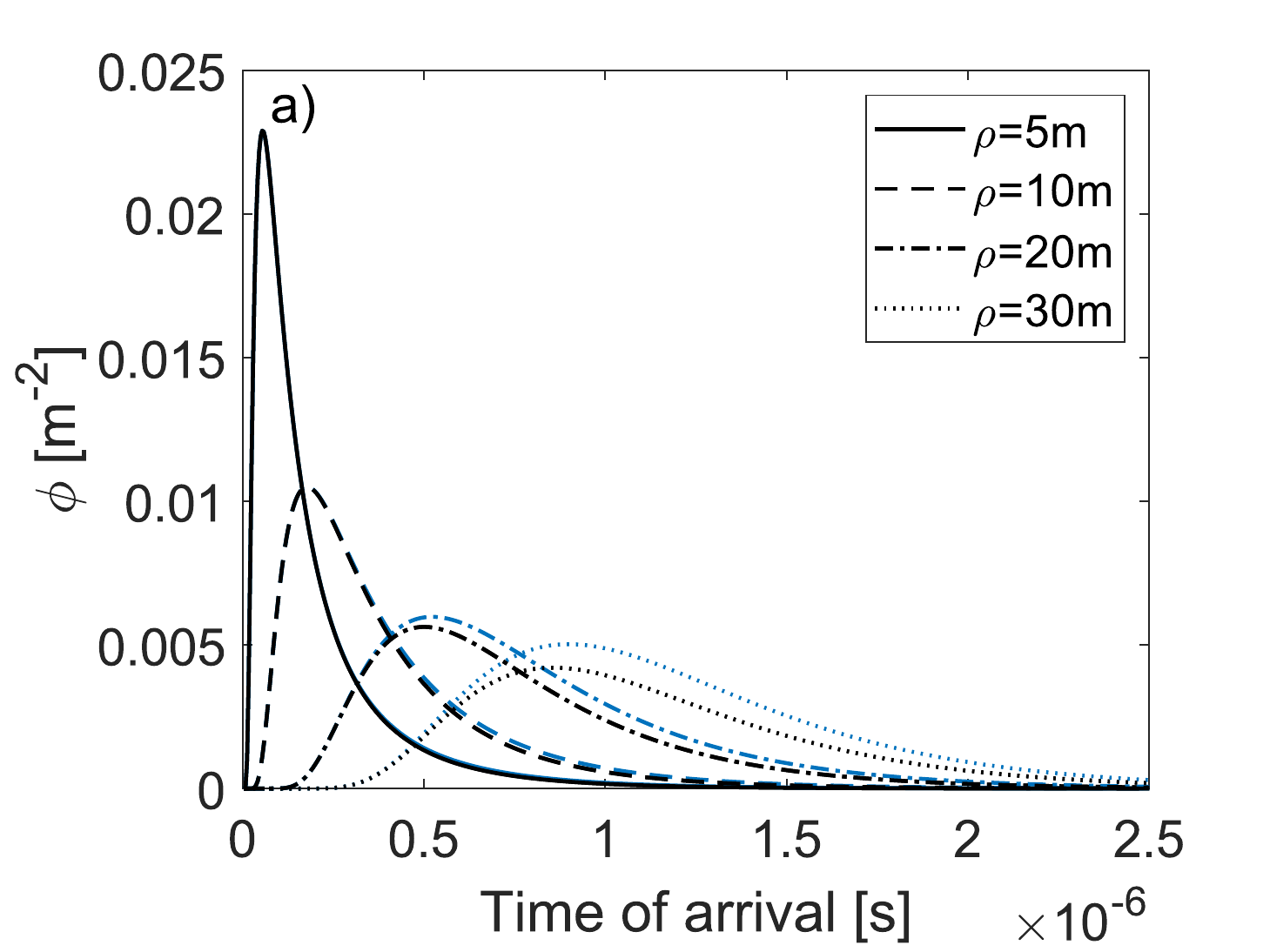}
    \includegraphics[width=0.49\textwidth]{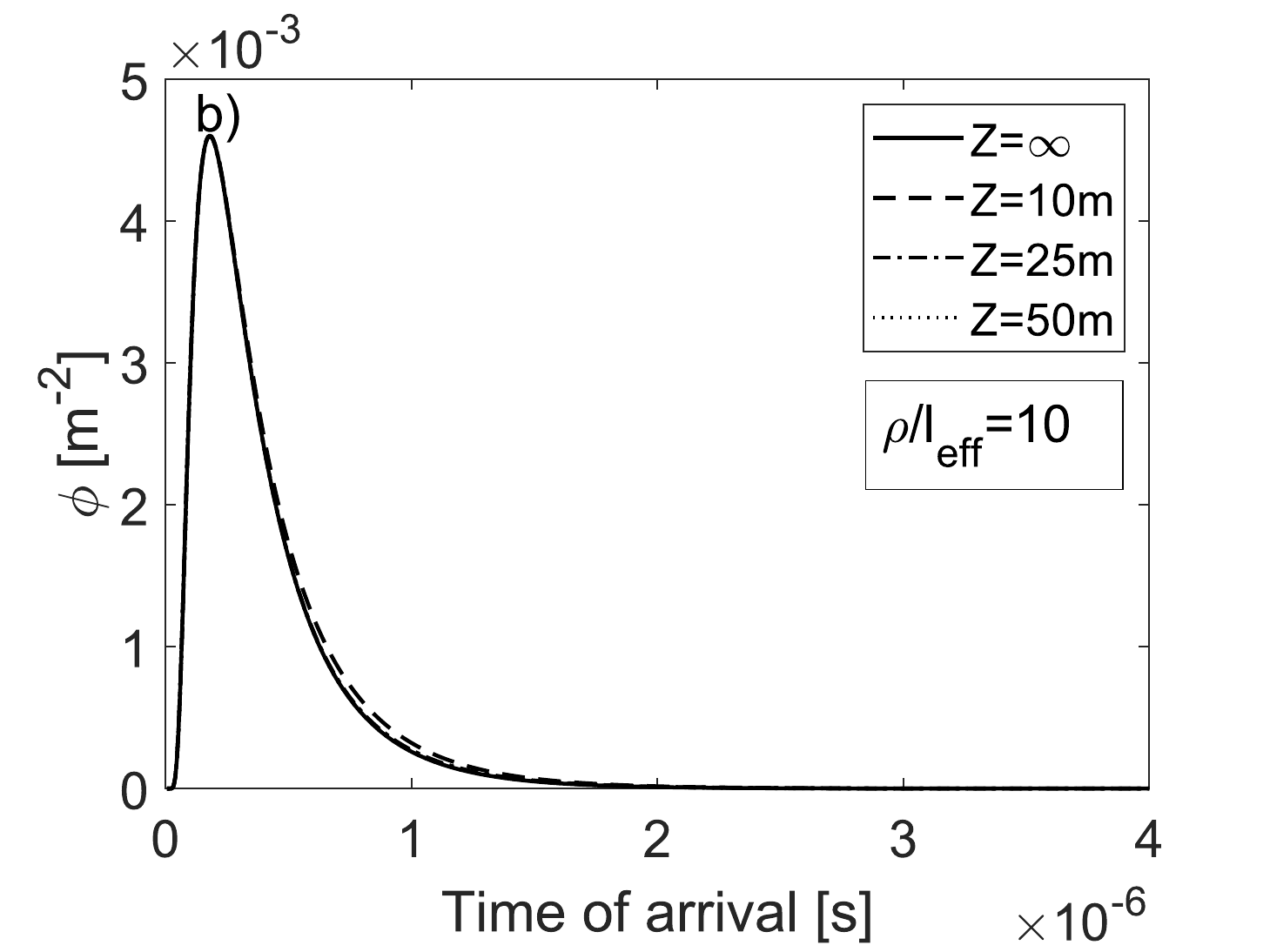}
    \includegraphics[width=0.49\textwidth]{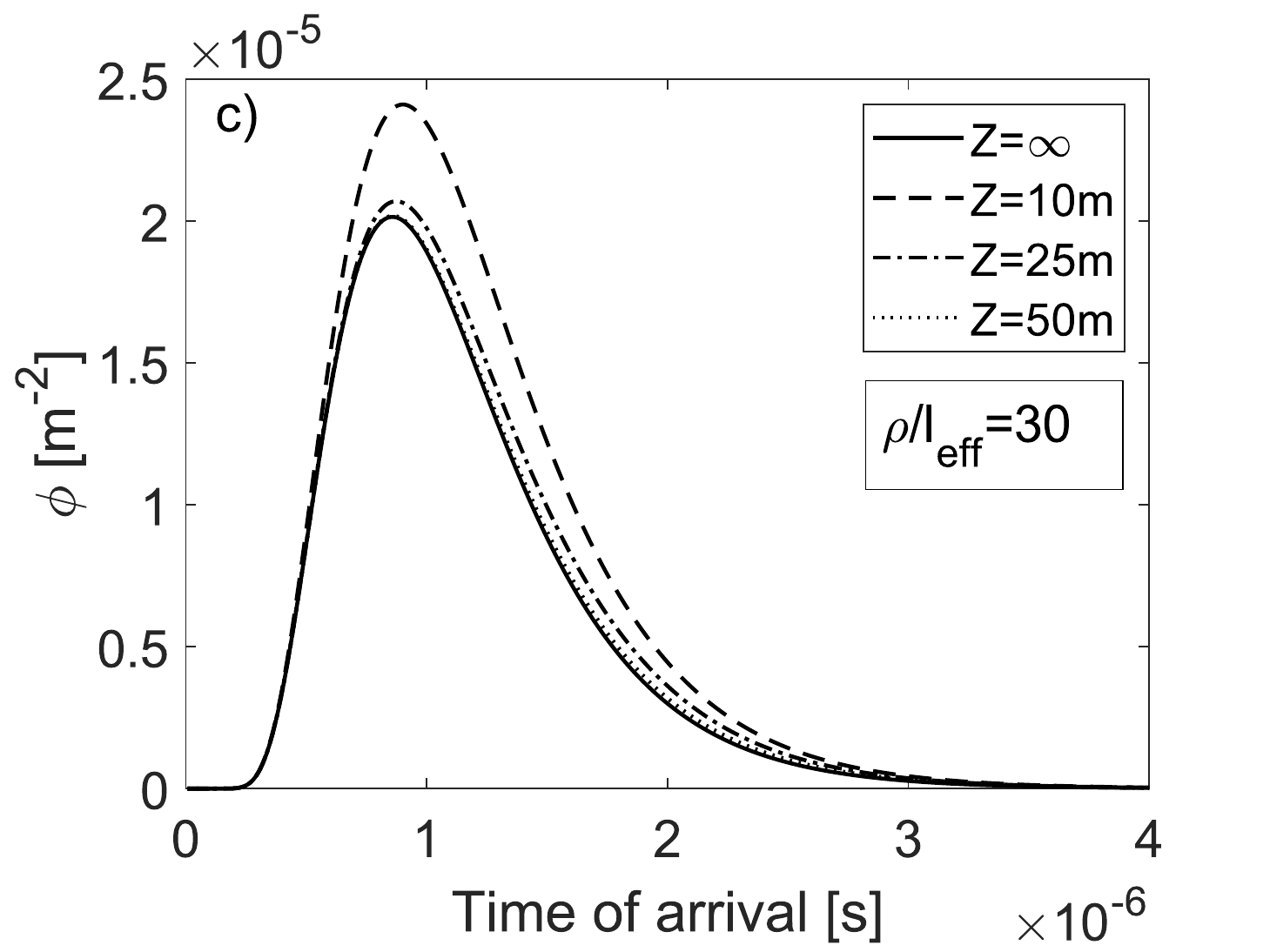}
    \includegraphics[width=0.49\textwidth]{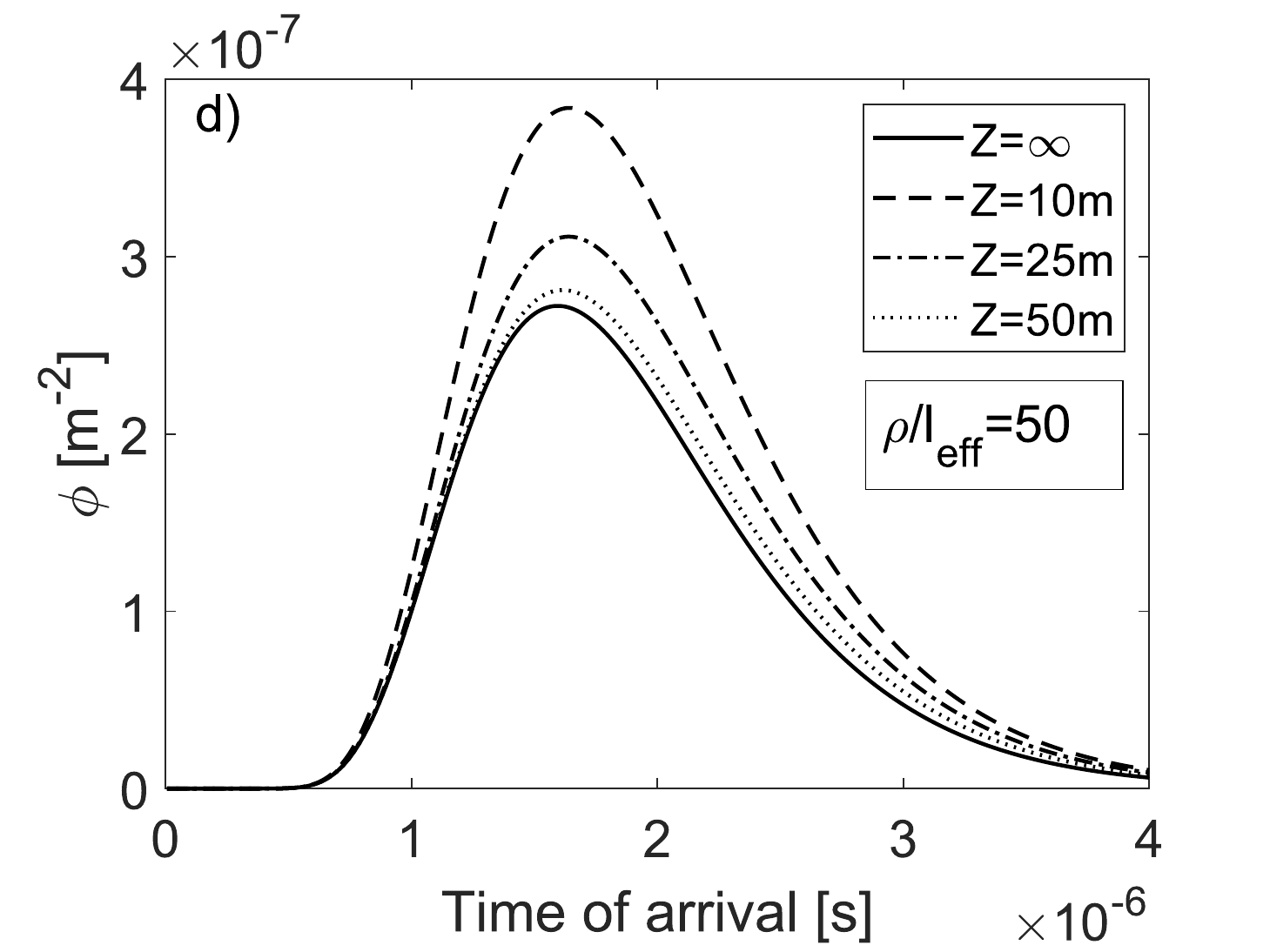}
    \caption{Time-dependent fluence rate for an instantaneous point-source and a detector separated by \(\rho\). (a) Different detector offsets, showing the fluence rate distribution for an infinitely deep slab (black lines) and with a bottom boundary 10m deep (blue line). (b-d) Distribution for different thicknesses, shown for detector separation of \(\rho\)=10m, 30m, 50m, respectively.}
    \label{fig:theo:tof}
\end{figure}

The above formula clearly shows the signature of three properties of the volume: diffusion constant, absorption constant, and thickness. Assuming the pure ice refractive index is known, the diffusion constant is directly related to only the scattering coefficient. The distribution \(\phi(\rho,t)\) on the surface can in principle be measured using photon counting detectors, which offer time-resolved detection of even dim light with resolution of the order of \(\sim100ps\). Since the diffusion approximation is only valid for many scattering steps, and the effective scattering length is of the order of \(\sim\)1m, it can be assumed that typical arrival times are at least of the order of many nanoseconds. Using these assumptions and a refractive index of \(1.31\) \cite{warren2019}, we can calculate the arrival time distributions at a hypothetical detector for different offsets of the detector from the light source, and for different thicknesses, shown in Figure \ref{fig:theo:tof}.

Figure \ref{fig:theo:tof}a shows the fluence rate as a function of time at four different detector positions for an infinitely deep volume. The blue curves are the corresponding fluence rates for a 10m-deep volume. The curves are all normalized to unit area for the infinite volume. It can be seen that the presence of a shallow boundary not only increases the fluence rate at the surface, but it also shifts the distribution to longer arrival times. Assuming a priori knowledge of optical properties, this is a clear signature of volume thickness that could potentially be used to invert a time-resolved measurement. With increasing distance between source and detector, this signature apparently grows larger as well. This can be understood in the picture of source and mirror source: Moving the detector further away from the source exponentially dampens the fluence rate from the source but does not proportionally change the separation from the mirror source at \(z=-2Z\), hence increasing the portion of the signal that bears the signature of volume thickness. In Figure \ref{fig:theo:tof}b-d we show the fluence rate distribution for different volume thickness in three different detector offsets, 10m, 30m and 50m respectively. Again, at large detector separation the difference between different volume thicknesses becomes more pronounced, but takes the shape of an almost symmetrical offset. With the parameters we assumed, it seems likely that thicknesses around 25m could be measured in principle. For larger thickness, a significant change in the shape of the curves is only present at very large distance from the source, where any remaining signal would be weak. It should be noted that normalizing the curves in Figure \ref{fig:theo:tof}a individually is not representative of an actual measurable signal, since absorption dominates the model behavior at larger source-detector separations, and actual measurable signal per injected pulse decreases. The magnitude of this decrease is better quantified in terms of time-integrated fluence rate. 

In addition to the signature of volume size there is a clear indication that the effective scattering and absorption coefficients can be extracted independently from a measured arrival time distribution, as was done for the infinite-volume situation \cite{price1997,ackermann2006}. Close examination of equation \ref{eq:theo:bounded} reveals that all terms present in the bounded volume model are of similar nature to the infinite volume solution from equation \ref{eq:rw:infinite}. Diffusive terms \(\sim \mathrm{exp}(-1/Dt)\) dampen short arrival times and bear the signature of the diffusion constant and hence the scattering coefficient. Absorptive terms \(\sim \mathrm{exp}(-\beta t)\) dampen dominantly long arrival times. We can therefore conclude that in a similar fashion to the deep measurements \cite{price1997,ackermann2006}, scattering and absorption coefficient can be independently extracted from measurements at the surface, assuming that the ice thickness is either known, or detector separation is small, in which case thickness has a minor influence on the detector signal. 

Another point worth noting is that the modification of the infinite solution achieved by introducing the mixed boundary condition at the surface is most prominent at small detector separation, mostly influencing the peak, and becomes negligible otherwise.

\subsection{Stationary sources and stationary measurement}

While time-resolved measurements have the potential of proving abundant information, they are not as simple as time-integrated intensity measurements. A stationary measurement, i.e., a continuous source and non-time-resolving detector that only measures intensity, would be much more desirable for a portable, low-cost measurement apparatus. Using non-time-resolved detection would yield the time-integrated, \textit{stationary} fluence rate, or short \textit{fluence} \(\phi_S(\rho)\) at the surface:

\begin{equation}
    \phi_S(\rho) = \int_0^\infty \phi(\rho,t,z=0)dt.
\end{equation}

\noindent Employing a stationary source gives an expression that can be shown to be identical with the stationary fluence. For this, the fluence rate at time \(t\) is the instantaneous fluence rate integrated over all past times from \(-\infty\):

\begin{equation}
    \int_{-\infty}^t\phi(\rho,t-t',z=0)dt' = -\int_{\infty}^0\phi(\rho,t,z=0)dt = \int_0^\infty \phi(\rho,t,z=0)dt = \phi_S(\rho).
    \label{eq:stationaryorstationary}
\end{equation}

\noindent We find that all time-dependencies in the fluence rate (equation \ref{eq:theo:bounded}) are of the form

\begin{equation}
    C \int_0^\infty dt \frac{1}{t^{3/2}} \mathrm{exp}\left(-\left(\frac{a}{t}+bt \right)  \right) = C \sqrt{\frac{\pi}{a}} \mathrm{exp}\left(-2\sqrt{ab}\right),
    \label{eq:integral}
\end{equation}

\noindent where \(C,a,b\) are constant. Calculating the fluence from equations \ref{eq:integral} for one term of the above form yields a stationary source term:

\begin{equation}
    \phi_{S+}(\rho) = \frac{1}{4\pi D} \frac{1}{\sqrt{\rho^2}}\mathrm{exp}\left(-\sqrt{\frac{3\rho^2}{l_{\mathrm{eff}}l_{\mathrm{abs}}}}\right).
\end{equation}

\noindent Applying this recipe to equation \ref{eq:theo:bounded} can be done term-by-term, yielding the complete fluence:

\begin{align}
    \phi_S(\rho) = &2\phi_{S+}(\rho^2+l_{\mathrm{eff}}^2) 
    + \phi_{S+}(\rho^2+(l_{\mathrm{eff}}+2Z)^2) 
    +\phi_{S+}(\rho^2+(-l_{\mathrm{eff}}+2Z)^2) \nonumber\\
    &+ \frac{2}{h}\int_0^\infty e^{-\frac{l}{h}}\phi_{S+}(\rho^2+(-l-l_{\mathrm{eff}})^2)dl 
    + \frac{2}{h}\int_0^\infty e^{-\frac{l}{h}}\phi_{S+}(\rho^2+(l+l_{\mathrm{eff}}+2Z)^2)dl.
    \label{eq:theosurf}
\end{align}

The result consists of the beforementioned 6 terms: 2 sources, 1 sink, and their mirror images, whereby the first two source terms are identical at \(z=0\). In terms of a injected energy, the time-integrated fluence rate has the units of \(\mathrm{[Jm^{-2}]}\). Integrating over a hypothetical detector aperture would yield the detected energy. Each term displays an exponential decay with separation \(\rho\) from the source, with each term again bearing different offsets to this value due to different effective source position. The exponentials fall off as \(\mathrm{exp}(-\sqrt{3\rho/l_{\mathrm{eff}}l_{\mathrm{abs}}})\). Hence, a fit to data would reveal the product of scattering and absorption coefficients. 

\begin{figure}
    \centering
    \includegraphics[width=0.49\textwidth]{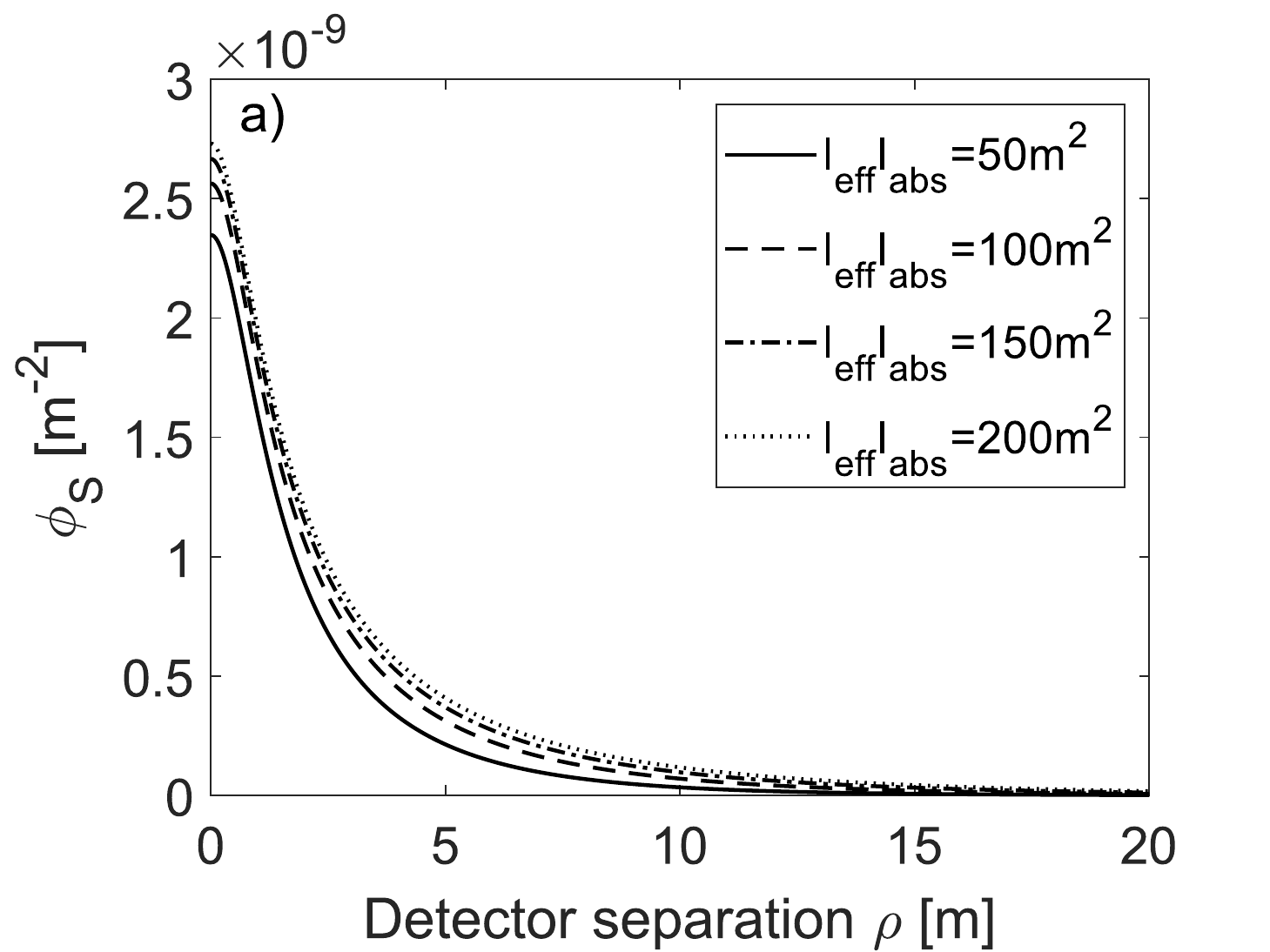}
    \includegraphics[width=0.49\textwidth]{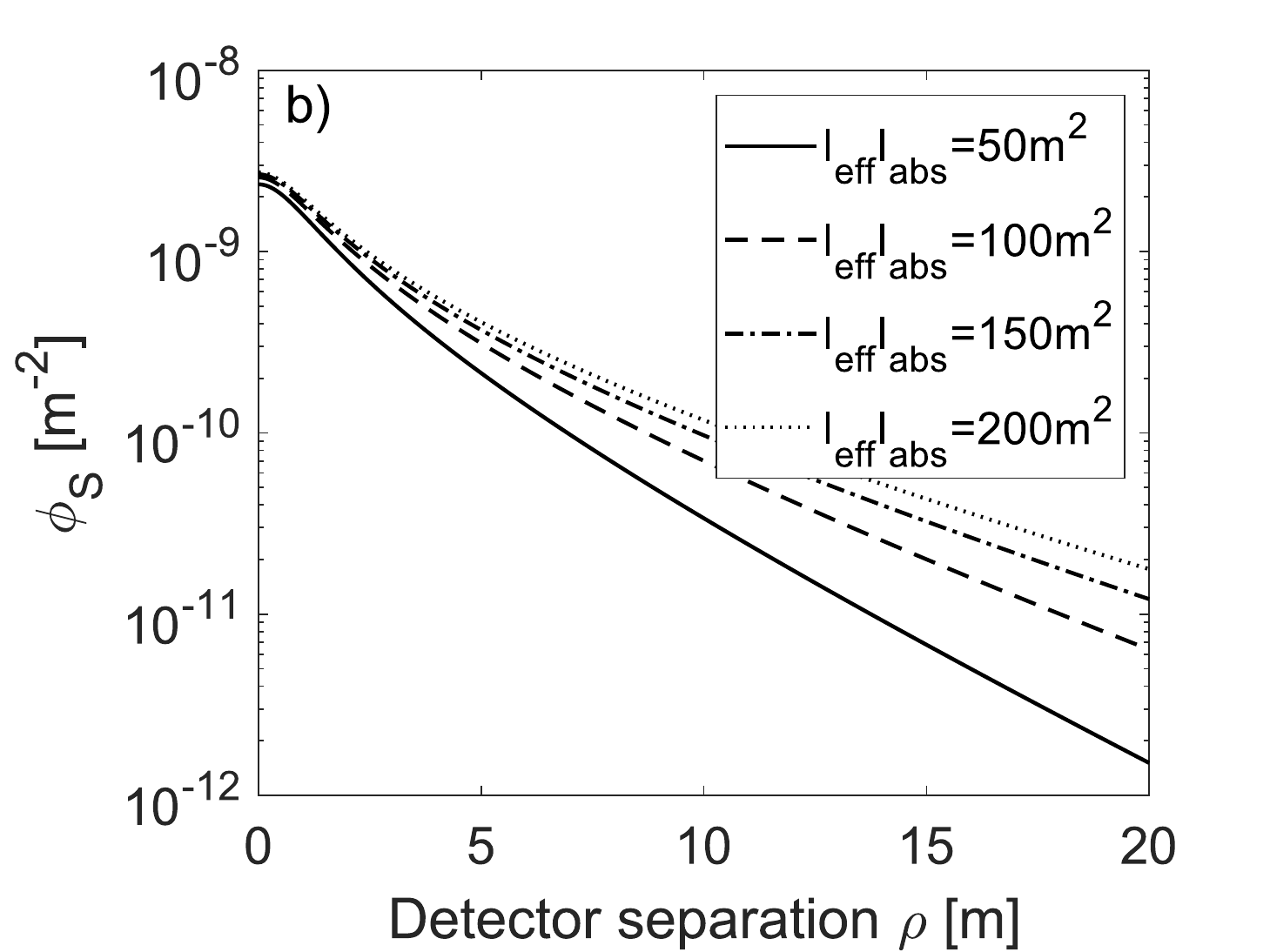}
    \includegraphics[width=0.49\textwidth]{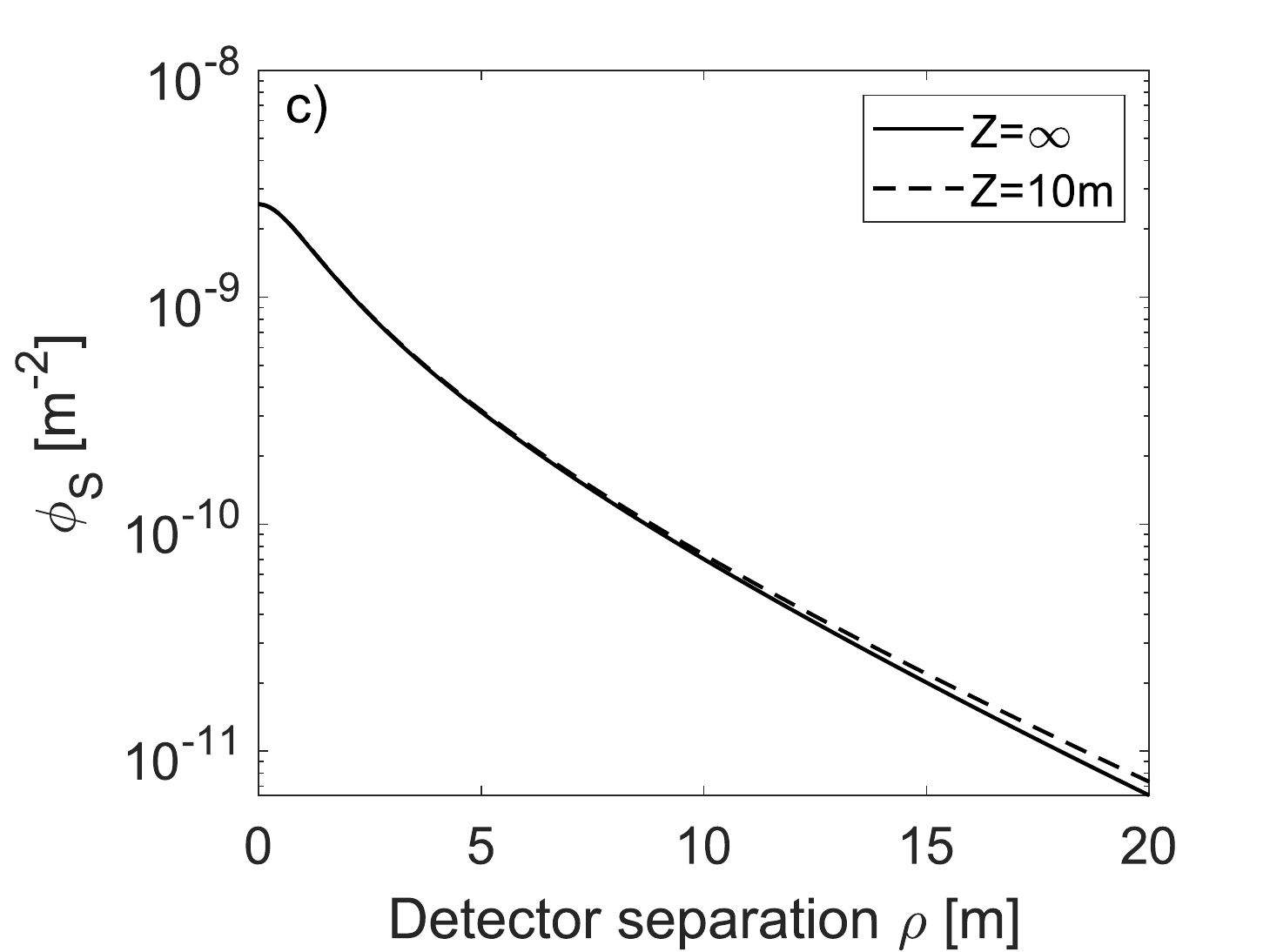}
    \includegraphics[width=0.49\textwidth]{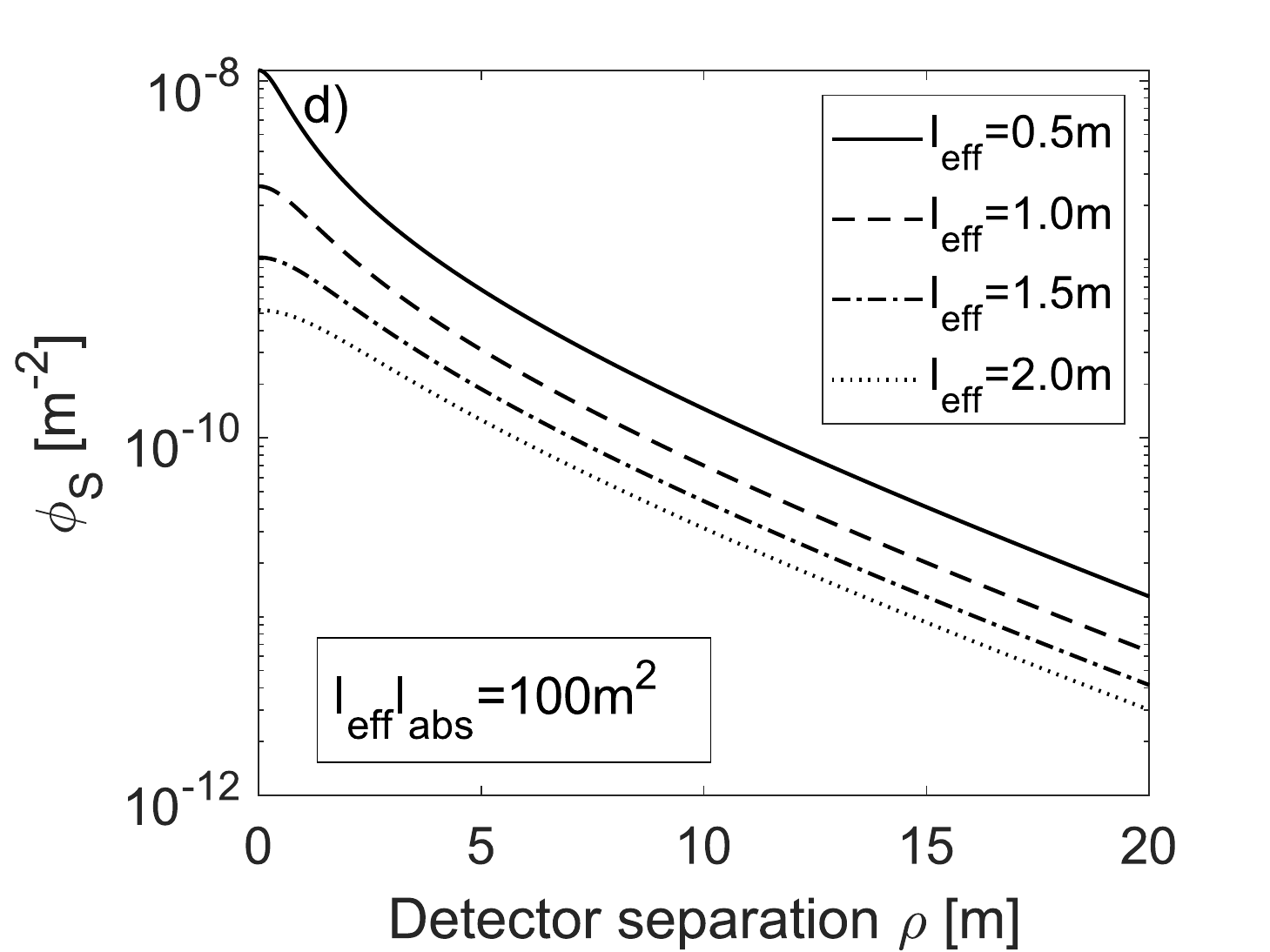}
    \caption{Time-integrated fluence rate for an instantaneous point-source and a detector separated by \(\rho\), compared for (a-b) different scaling parameters \(l_{\mathrm{eff}}l_{\mathrm{abs}}\) on linear and log-scale, respectively, (c) infinite and 10m-deep volume, and (d) for constant scaling constant \(l_{\mathrm{eff}}l_{\mathrm{abs}}\) but varying scattering length \(l_{\mathrm{eff}}\).}
    \label{fig:theo:stat}
\end{figure}

The integrated fluence rate for different system parameters is shown in Figure \ref{fig:theo:stat}. Panel a shows the fluence rate for  constant scattering length \(l_{\mathrm{eff}}=1\)\,m, but different absorption lengths, resulting in a different scaling parameter \(l_{\mathrm{eff}}l_{\mathrm{abs}}\). Panel b shows the same data on a log-scale. The difference is most apparent in the tails of the distribution, where the curves cover about one order of magnitude for a realistic range of parameters, displaying different slopes on a log-scale.

Figure \ref{fig:theo:stat}c shows the same curve for two different geometries, the infinitely deep volume, and a shallow one with a depth of 10m. It is clear that the difference is minor and only shows up in the tail of the distribution. This hints that the integrated fluence rate is not an ideal measure of volume geometry, especially compared to the time-dependent fluence rate. This highlights that stationary measurements are more useful for extracting optical properties rather than measuring thickness - especially so since the measurement is largely independent of thickness.

Figure \ref{fig:theo:stat}d shows the distribution for different scattering lengths, but constant overall scaling constant \(l_{\mathrm{eff}}l_{\mathrm{abs}}\). In comparison to panel b the slope in the tail (corresponding to the exponential term) is retained, while the behavior at small detector separation is different for each set of parameters. A difference is only visible for small detector separation, where the diffusion approximation may not hold. This implies that independent measurement of scattering and absorption is not possible in this configuration.

Lastly, the integrated fluence rate can answer question of the absolute measurable signal level in a time-resolved measurement, and its fall-off with detector position \(\rho\). According to equation \ref{eq:stationaryorstationary}, the integrated fluence rate is identical to the integral over the entire arrival time distribution shown in Figure \ref{fig:theo:tof}. Therefore, the integrated fluence rate is a measure for the total measured signal at position \(\rho\) in a time-resolved measurement, revealing the overall system loss.

\section{Monte Carlo Simulations}

The model developed so far is based on several assumptions and approximations. We're assuming that the physical Mie scattering process can be approximated with an effective isotropic process, and that the diffusion approximation holds, specifically at the boundaries. The bottom boundary is modeled as a perfect mirror, which surely is far from reality. To test if these assumptions are reasonable, we are comparing the model to Monte Carlo simulations implementing a more physical process. However, these simulations have their own limitations, especially since a computationally efficient physical scattering phase function has to be chosen. Agreement with a realistic, physical scattering process is only guaranteed if the phase function is well known.

The simulations calculate the trajectory of an individual photon. Scattering is implemented using a Henyey-Greenstein distribution \cite{toublanc96} around the pitch-axis of the photon flight vector, and a uniform distribution around the roll axis, which is the vector itself, using the Rodrigues' rotation formula to calculate. The asymmetry factor for the employed Henyey-Greenstein distribution is \(g=\langle \cos \theta \rangle=0.75\). The two random angles \(\theta,\gamma\) in each scattering step are obtained from uniformly distributed random numbers \(\xi_1,\xi_2\in\{0,1\}\) using the inverse cumulative Henyey-Greenstein density function

\begin{align}
    &\theta = \mathrm{cos}^{-1} \left[\frac{1}{2g} \left(1 + g^2 - \left(\frac{1-g^2}{1+g-2g\xi_1}\right)^2 \right)\right], \\
    &\gamma = 2\pi\xi_2 .
\end{align}

Since this approximation to the true Mie scattering distribution only accounts for the forward component of scattering, artifacts around the injection point have to be expected. While the bottom boundary is assumed to be lossless and cosine-distributed

\begin{align}
    &\theta = \mathrm{sin}^{-1}\left(\xi_1\right), \\
    &\gamma = 2\pi\xi_2 .
\end{align}

Given that the bedrock of a glacier will have some degree of microscopic and macroscopic roughness, this approximation seems appropriate, and is in line with experimentally observed emission profiles \cite{watson1971}. In the presence of a water or air layer below the ice, a (partial) specular reflection may be more appropriate. For cases \(l_{eff}\ll Z\), the distribution will not matter, since randomization of the distribution will occur regardless of the scattering distribution at the interface. For all cases under investigation in the following chapter we observe no significant differences in the simulated distributions for reflective, cosine or uniformly distributed angles at the lower interface.

\begin{figure}
    \centering
    \includegraphics[width=0.49\textwidth]{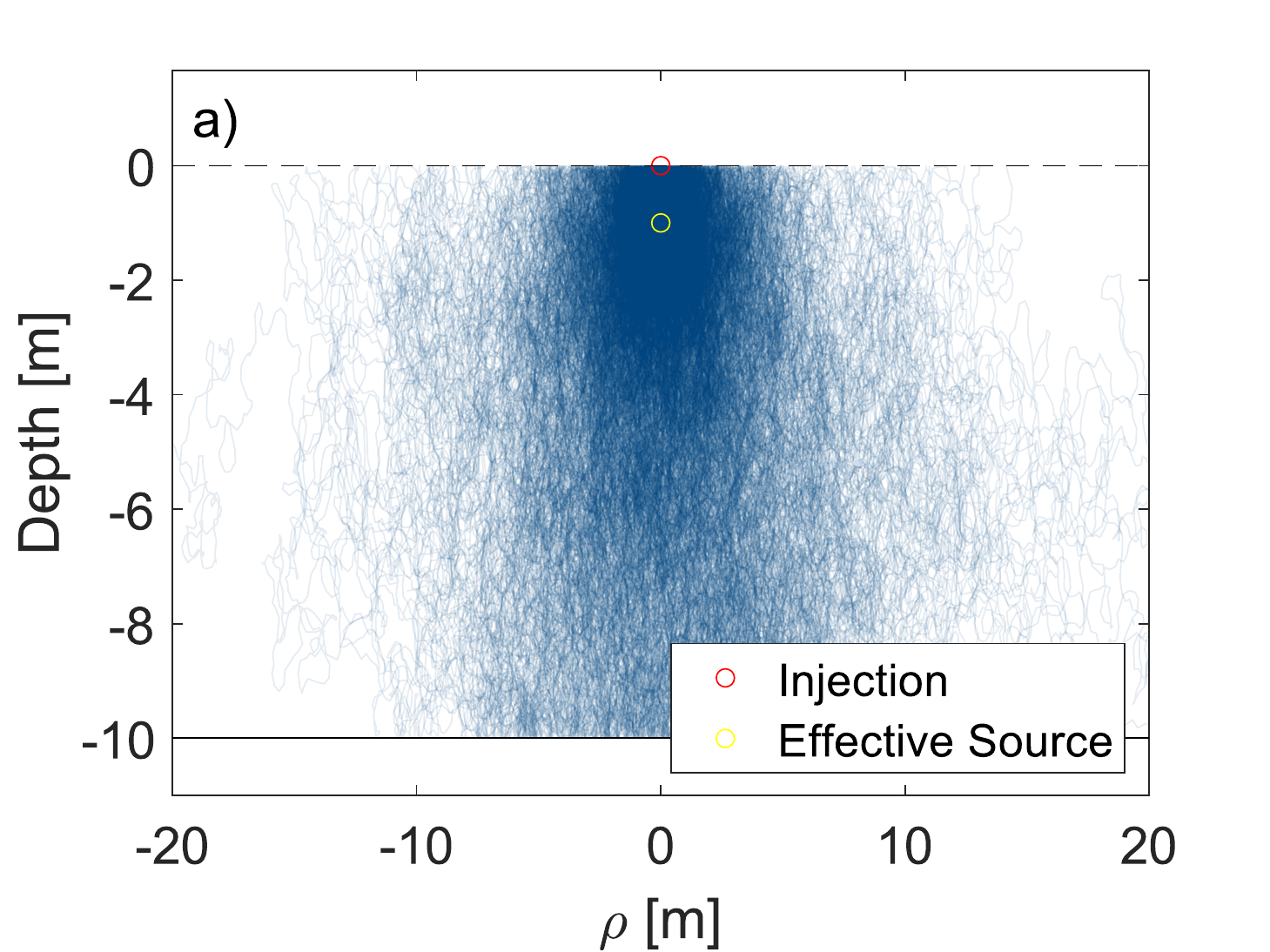}
    \includegraphics[width=0.49\textwidth]{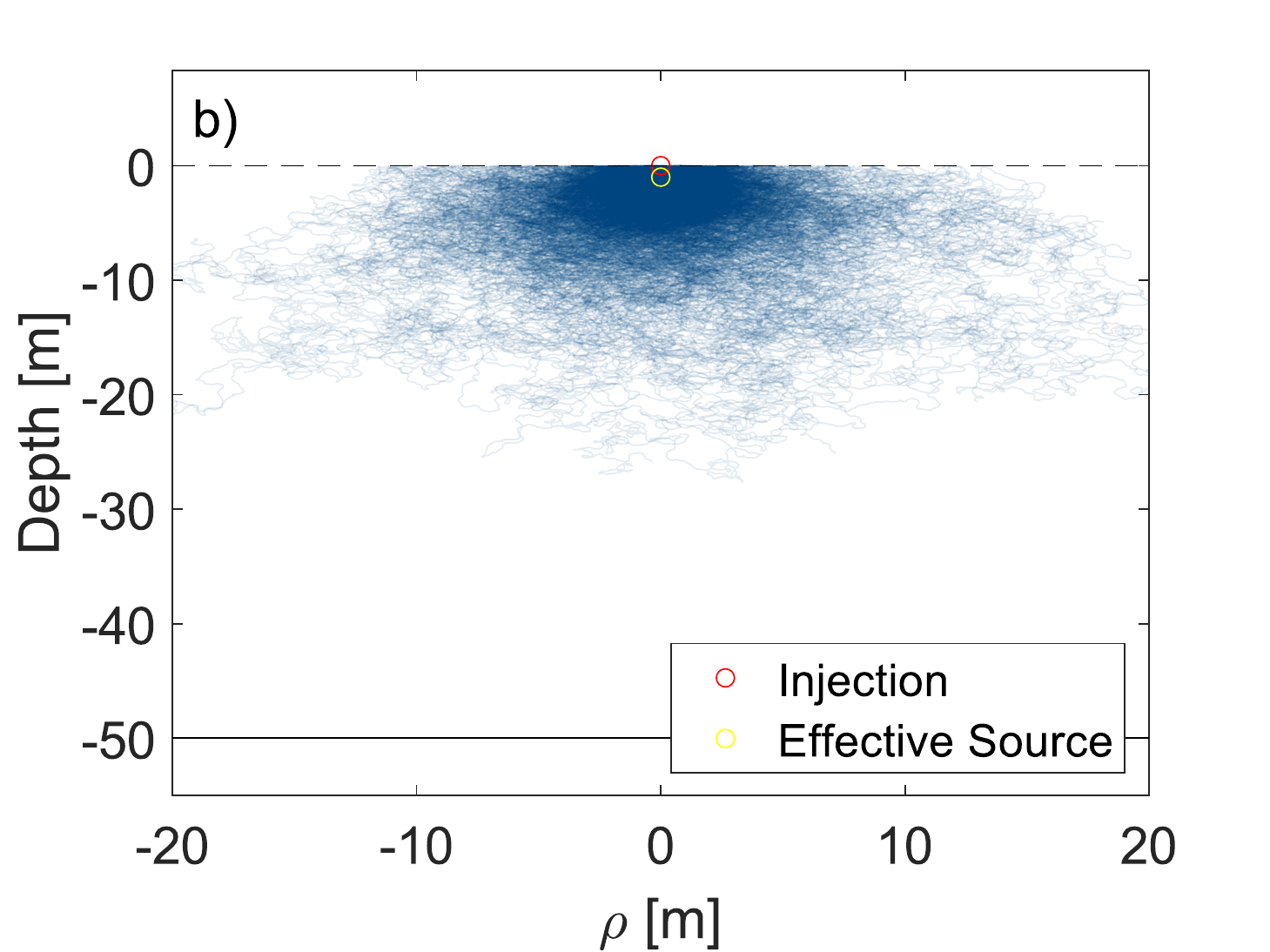}
    \caption{Simulated random flights for 2000 photons with constant scattering length \(l_{\mathrm{eff}}=1\)m and absorption length \(l_{\mathrm{abs}}=100\)m. The bottom boundary implements a total, diffuse (cosine-distributed) reflection, the top boundary implements Fresnel reflection. Light is injected straight down, getting scattered after one physical scattering length, whereas the model's effective source would be one effective scattering length down. (a) Shallow volume,  \(Z=10\)m, (b) Deep volume, \(Z=50\)m.}
    \label{fig:simu1}
\end{figure}

A Fresnel reflection is implemented at the top boundary, assuming unpolarized light and summing the contribution of both linear polarizations, assuming a refractive index of ice of 1.31 \cite{warren2019}. For this refractive index, total internal reflection takes place for trajectories that intersect the surface at an angle larger than \(\theta_C=50^\circ\). Each step implements absorption in the form of a survival probability of \(\mathrm{exp}(-l_{\mathrm{sca}}/l_{\mathrm{abs}})\) and trajectories are terminated either by absorption or return to the surface. Figure \ref{fig:simu1} shows 2000 trajectories for a deep and a shallow volume and denotes injection point, effective source position and boundaries. The overlayed trajectories display the usual pear-shaped distribution.

For each photon trajectory the endpoint and time-of-flight are accessible. A subset of trajectories ending in a defined detector area of 0.01\(\mathrm{m^2}\) - achievable with off-the-shelf optics - is evaluated in histograms to simulate the outcome of a time-of-flight measurement. Surface intensity measurement can be simulated easily by counting all trajectories ending within detector areas along a line on the surface.

\section{Feasibility of measurements}

The simulations can be compared to the theoretical model. The simulations can easily predict photon count rates and measurement duration for a given source brightness. For a successful measurement of any parameters, it should be possible to extract these parameters by fitting the theoretical curves to the simulated data.

\subsection{Intensity measurements}

\begin{figure}
    \centering
    \includegraphics[width=0.49\textwidth]{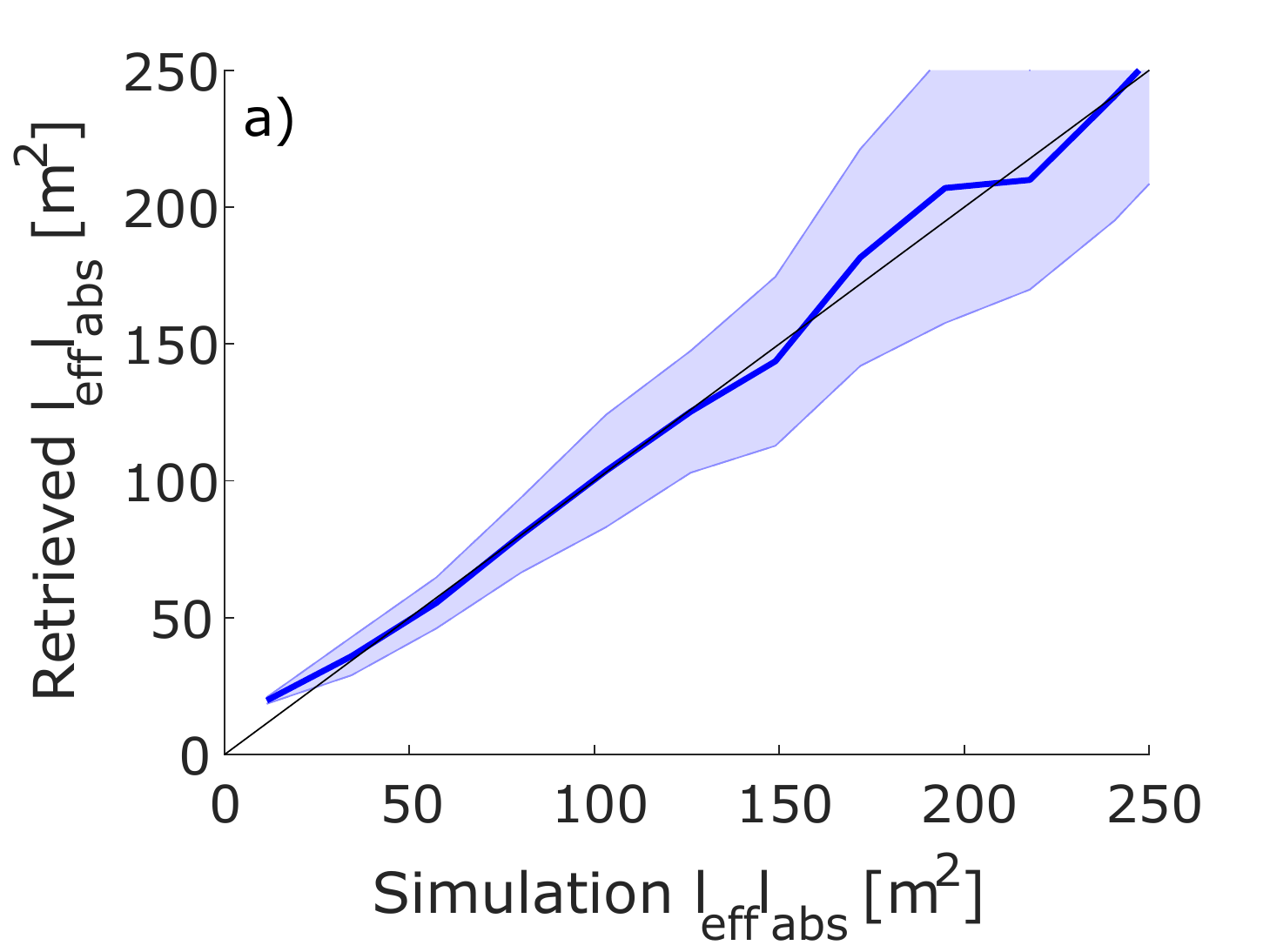}
    \includegraphics[width=0.49\textwidth]{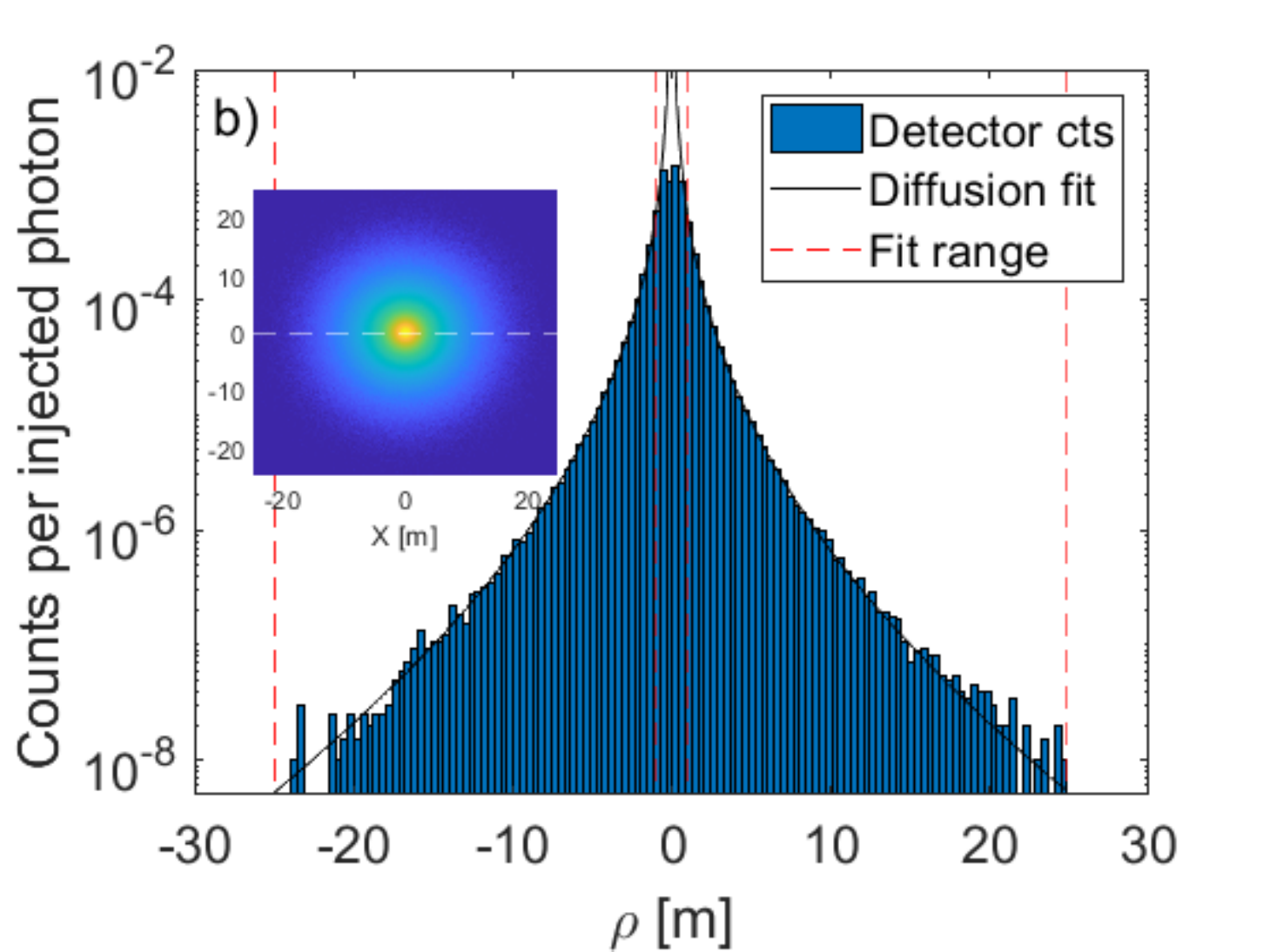}
    \caption{Simulated time-integrated detector counts per injected photon on the surface for 100cm\(^2\) detector area with \(10^8\) simulated photon trajectories. (a) Reconstructed vs simulated scaling constant (blue curve) over the ideal ratio of 1 (black curve). Points shown contain combination of scattering and absorption lengths in the range of \(l_{\mathrm{eff}}=0.5...1.5m\) and \(l_{\mathrm{abs}}=20...250m\). Data is averaged in bins of width \(20m^2\) with \(R^2=0.99\) (\(R^2=0.93\) for the raw data). (b) Simulation with scaling constant \(l_{\mathrm{eff}}l_{\mathrm{abs}}=130\mathrm{m^2}\) and \(10^8\) photon trajectories. The solid line is a fit of the mixed-boundary diffusion model to the simulated histogram, excluding the center portion. Reconstructed scaling constant is \(l_{\mathrm{eff}}l_{\mathrm{abs}}=134(\pm7)\mathrm{m^2}\). The inset shows the two-dimensional distribution on the surface with the dashed line indicating where the cut was taken.}
    \label{fig:simusurf1}
\end{figure}

The easiest measurement to implement is an intensity measurement as a function of detector position. Here, photon counting detectors such as photo multiplier tubes or avalanche photo diodes offer good efficiency, temporal resolution, and more importantly a dynamic range over many orders of magnitude. This helps when extracting meaningful data from measurements. To ascertain how well this technique would fare given the complicated nature of equation \ref{eq:theosurf}, we performed simulations for different combinations of parameters \(l_{\mathrm{eff}}\) and \(l_{\mathrm{abs}}\), corresponding to a scaling parameter \(l_{\mathrm{eff}}l_{\mathrm{abs}}\) between 20\(\mathrm{m^2}\) and 250\(\mathrm{m^2}\), covering a wide range of expected values. The results are shown in Figure \ref{fig:simusurf1}.

For fitting the theoretical model, we scale it by detector area. Simulated data points \(y\) are inversely weighted as \(1/y\) to account for the large dynamic range of 4-5 orders of magnitude. As a result, the scaling parameter used for the simulation can be extracted. Accuracy is in the range of 10\% for smaller values. Faithful fitting seems increasingly difficult at larger values for \(l_{\mathrm{eff}}l_{\mathrm{abs}}\). While scattering and absorption length are used as independent fitting parameters as prescribed by equation \ref{eq:theosurf}, only their product yields reasonable agreement with the expected value. We believe that this is caused by the invalidity of the diffusion approximation in proximity to the detector. In Figure \ref{fig:simusurf1}b we show one example of simulated measurement data for a scaling constant of \(l_{\mathrm{eff}}l_{\mathrm{abs}}=130\mathrm{m^2}\). The fit retrieves a value of \(l_{\mathrm{eff}}l_{\mathrm{abs}}=134(\pm7)\mathrm{m^2}\). In the lowest counting detector positions, detection probability per injected photon is of the order of \(10^{-9}\). Assuming injection rates of \(10^{15}\mathrm{s}^{-1}\), which corresponds to roughly \(1\)\,mW of optical power, we can expect count rates of the order of \(10^6s^{-1}\). This is equivalent of about \(1\)\,pW of optical power. Accounting for limited detector and collection efficiency, these estimates leave ample room to clear ambient nighttime background using photon counting detectors or highly efficient photo diodes.

Note that product of absorption and scattering coefficient is sufficient above 600\,nm, where absorption is well known and independent of impurities \cite{warren2006,cooper2020}. Physical scattering length \(l_{\mathrm{sca}}\) has to be independent of wavelength, and the wavelength-dependence of the scattering angle distribution (asymmetry factor) can be modeled through Mie scattering theory, which can serve as a waypoint to extrapolate scattering coefficients measured above 600\,nm to short wavelengths.

Even though information for large scaling parameters is hard to retrieve with the presented method, the small experimental effort necessary makes it an attractive measurement regardless of its limitations. Only a diode laser source, detector and collection optics are required.

\subsection{Time-of-flight measurements}

Time-resolved measurements can potentially provide more information, but at the cost of lower counting rates, since the signal will be distributed in many discrete time bins. We first assume that the volume depth is either known or can be assumed sufficiently large not to play a role. Using these assumptions, we apply a least-square fit to the simulated arrival time histograms and attempt to retrieve the simulation parameters for scattering and absorption length. 

\begin{figure}
    \centering
    \includegraphics[width=0.49\textwidth]{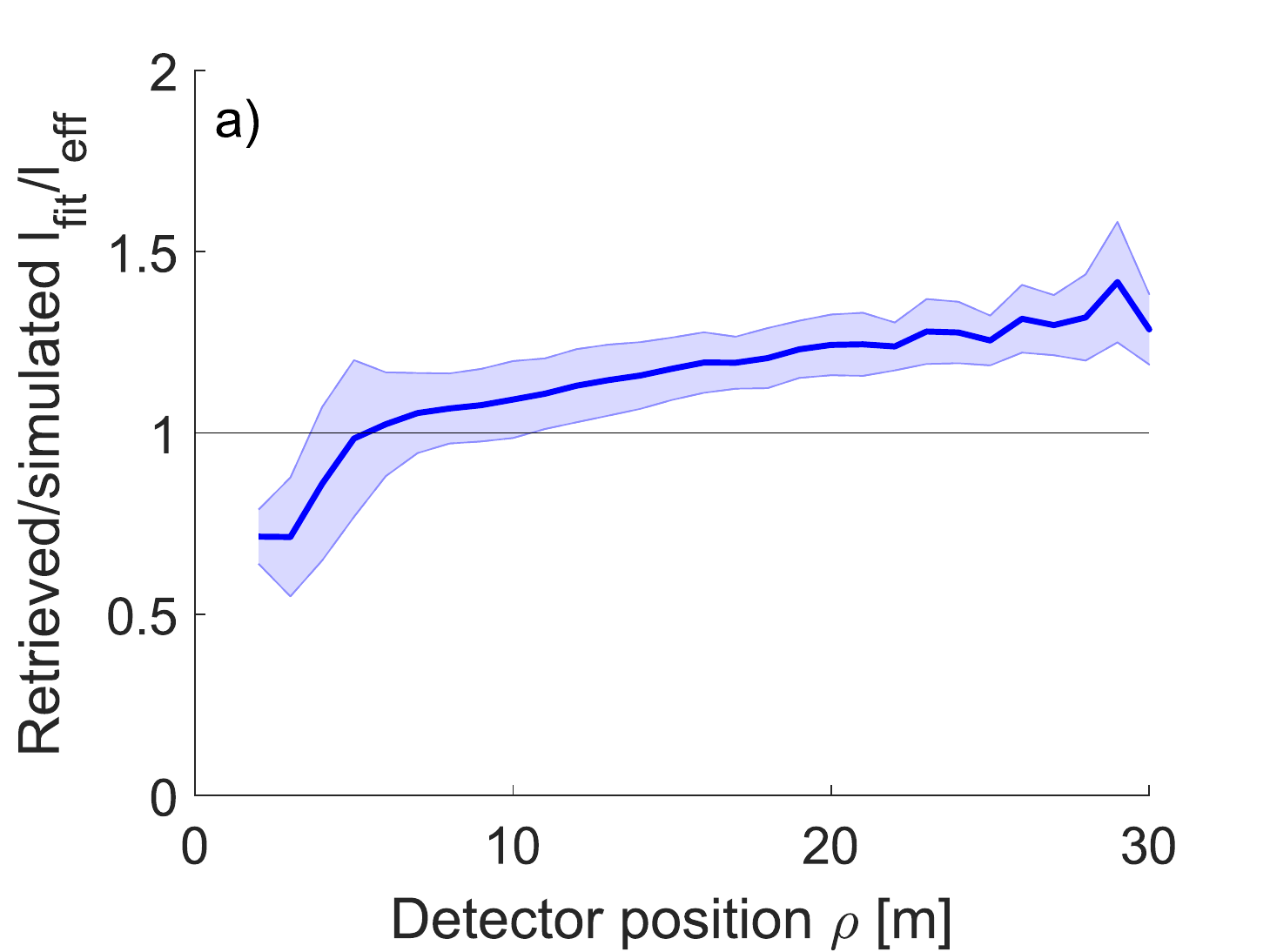}
    \includegraphics[width=0.49\textwidth]{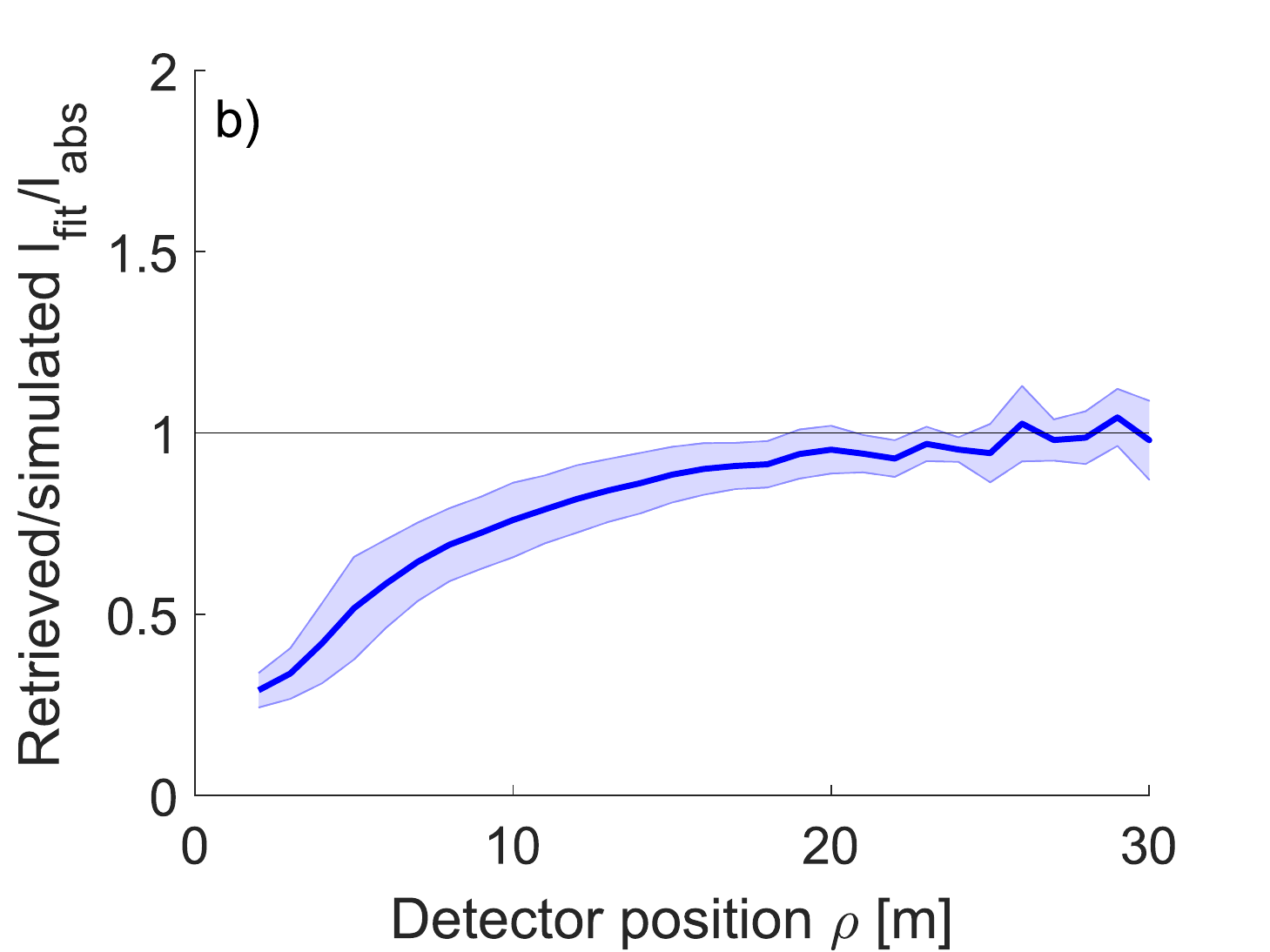}
    \includegraphics[width=0.49\textwidth]{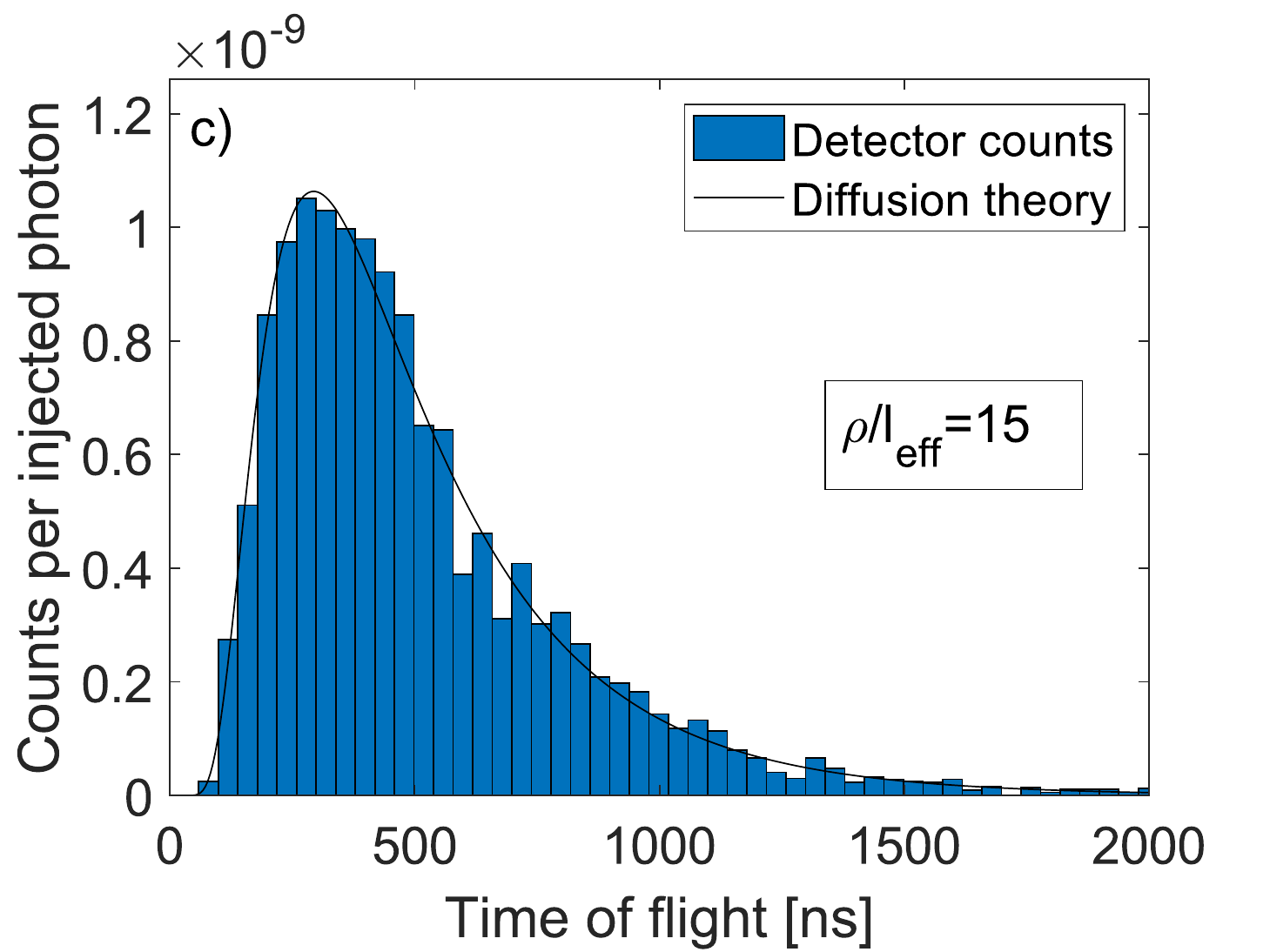}
    \caption{Simulated arrival time histogram of detector counts per injected photon for 100cm\(^2\) detector area, fitted with diffusion theory curves. The dataset combines nine different combinations of scattering and absorption length in the range of \(l_{\mathrm{eff}}=0.5...1.5m\) and \(l_{\mathrm{abs}}=25...250m\). (a) Reconstructed vs simulated effective scattering length. (b) Reconstructed vs simulated absorption length. (c) Example on a simulated arrival time histogram at a detector position 15m away from the source, scattering length of \(l_{\mathrm{eff}}=1\)m and absorption length \(l_{\mathrm{abs}}=100\)m. The least-square fit retrieves parameters \(l_{\mathrm{eff,fit}}=1.14(\pm0.03)\)m and \(l_{\mathrm{abs,fit}}=88(\pm5)\)m.}
    \label{fig:simuhist}
\end{figure}

The results for several combinations of scattering and absorption length are shown in Figure \ref{fig:simuhist}. The ratio between retrieved and simulated parameter is shown as a function of detector position. It can be seen that for large detector offset, both parameters asymptotically reproduce the simulated parameters within the variance of the dataset. Achievable relative error is below 20\% for detector positions at least 15m away from the source. The asymptotic behavior can be explained by the fact that the diffusion approximation only holds for large detector separations. This appears to be much more apparent for time-resolved measurements than it is for time-integrated ones. Figure \ref{fig:simuhist}c shows an example of a fit. In general, these fits are much more robust than the ones of surface distribution. This is due to the separate influence of scattering and absorption length on different sections of the curve, but also due to the smaller dynamic range of 1-2 orders of magnitude. We found that fit solutions are largely independent of starting values.

For the evaluation of the simulated arrival time distributions we chose a bin width of 50\,ns, which yields a reasonable compromise between sampling the distribution with enough points and producing sufficient number of counts per bin. Using this temporal resolution of 50\,ns we estimate return loss of the order of \(10^{-10}\) for the bins in the tail of the distribution. This is well above the temporal resolution of modern photon counting detection systems, which is typically of the order of 100ps for combined response of the detector and electronics. Since optical parameters can be retrieved using such bin widths, the use of inexpensive nanosecond-pulsed lasers is possible. These lasers typically provide pulse energies of the order of 1nJ (\(10^9\) photons per pulse), with pulse repetition rates of \(10^5\mathrm{s}^{-1}\), limited by the width of the arrival time distribution. Taking combined detector and collection efficiency of \(10^{-3}\) into account, we expect count rates for the lowest counting bins of the order of \(10s^{-1}\). This is reasonable for a time-resolved measurement, where any background is constant. It is worth noting that the technique gives a truly independent measurement of scattering and absorption length, without assumptions or reference values used previously \cite{warren2006,cooper2020}

\subsection{Thickness}

\begin{figure}
    \centering
    \includegraphics[width=0.49\textwidth]{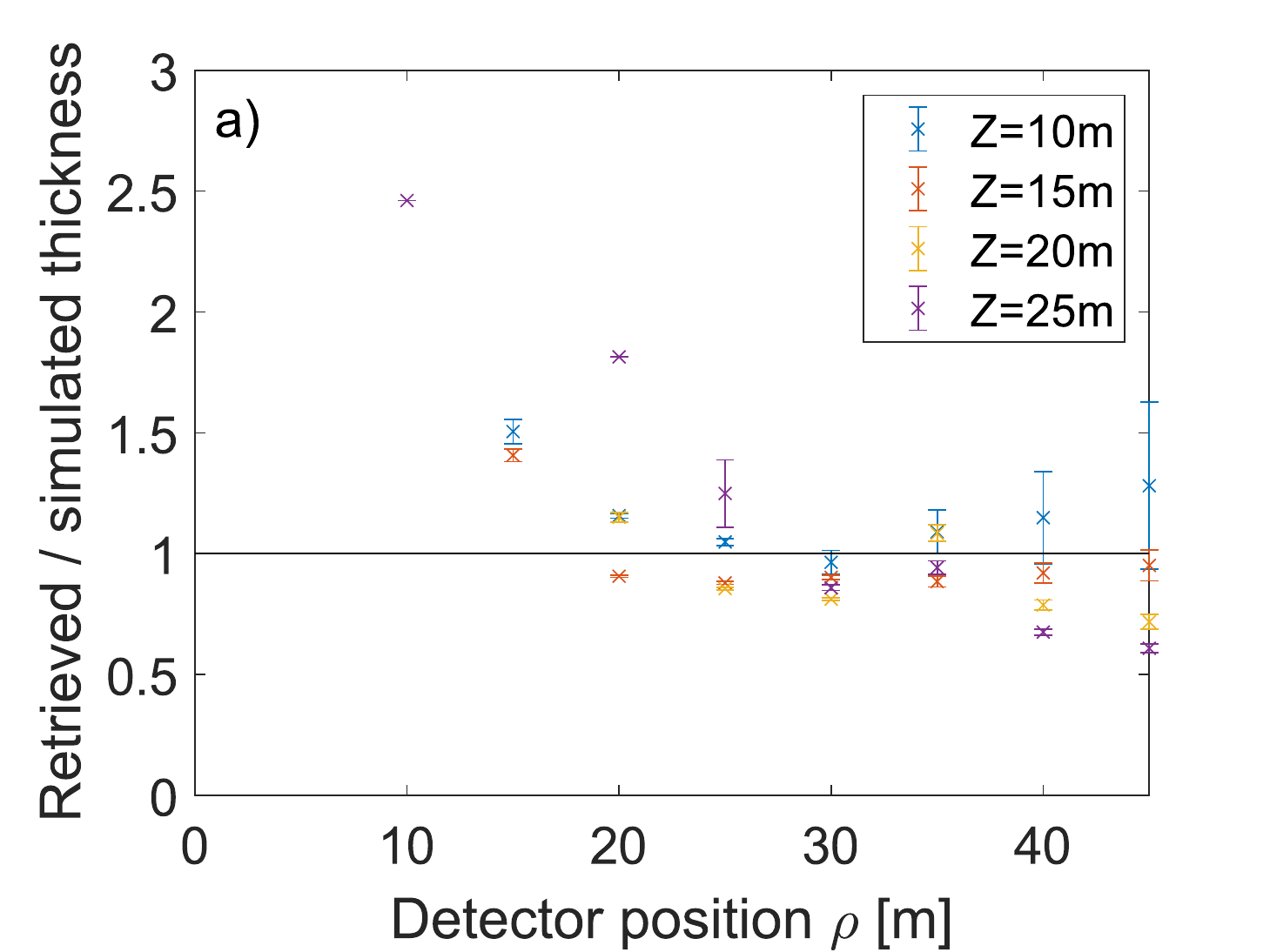}
    \includegraphics[width=0.49\textwidth]{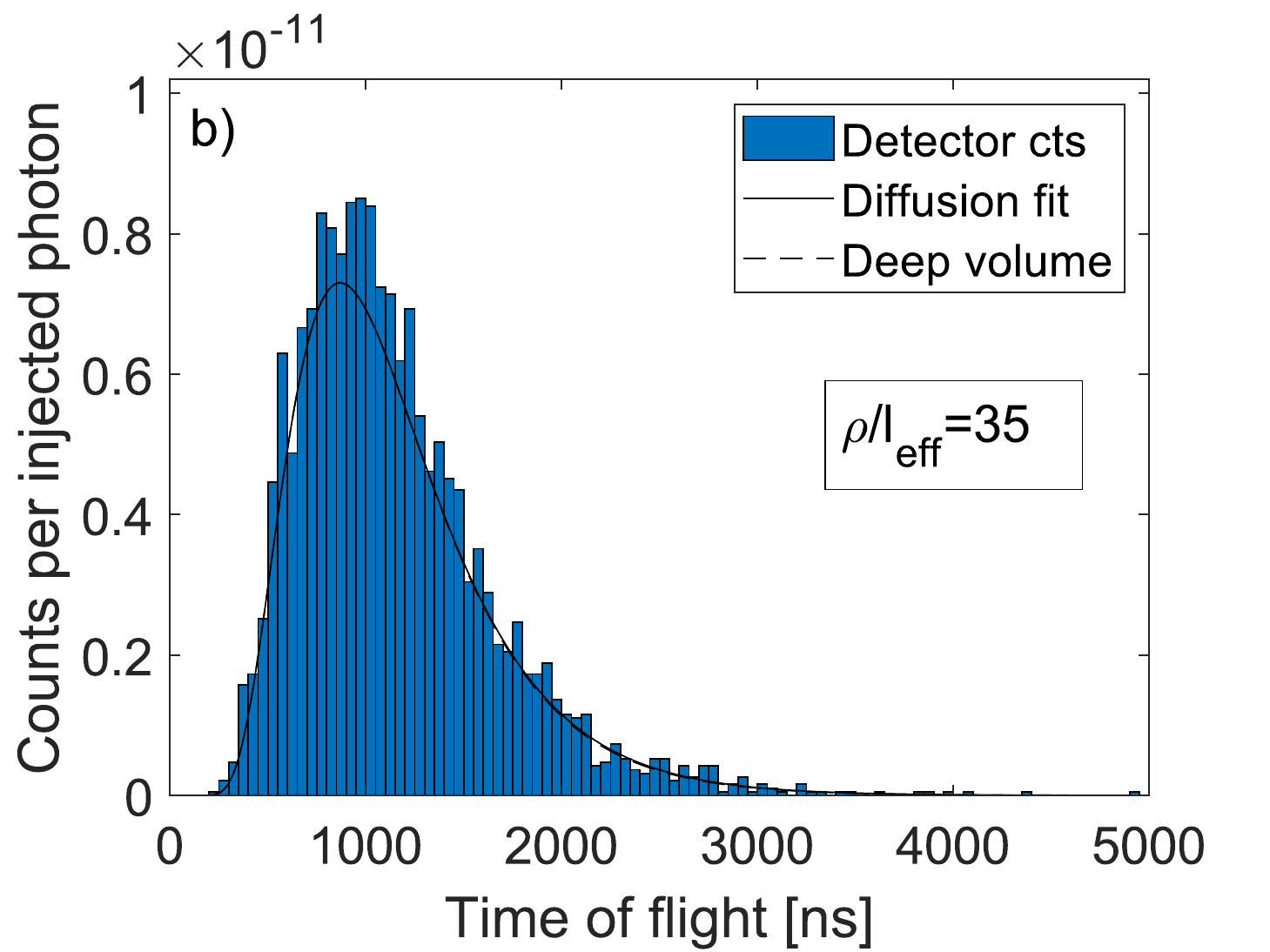}
    \caption{Simulated arrival time histogram of detector counts per injected photon for 100cm\(^2\) detector area. (a) Reconstructed vs simulated volume depth. Confidence intervals for unreliable fits  are not shown. (b) Example on a simulated arrival time histogram for a volume depth of \(Z=25m\) with the detector situated at \(\rho=35\)m. The recovered volume depth is \(Z=24(\pm9)m\).}
    \label{fig:simudepth}
\end{figure}

\begin{figure}
    \centering
    \includegraphics[width=0.49\textwidth]{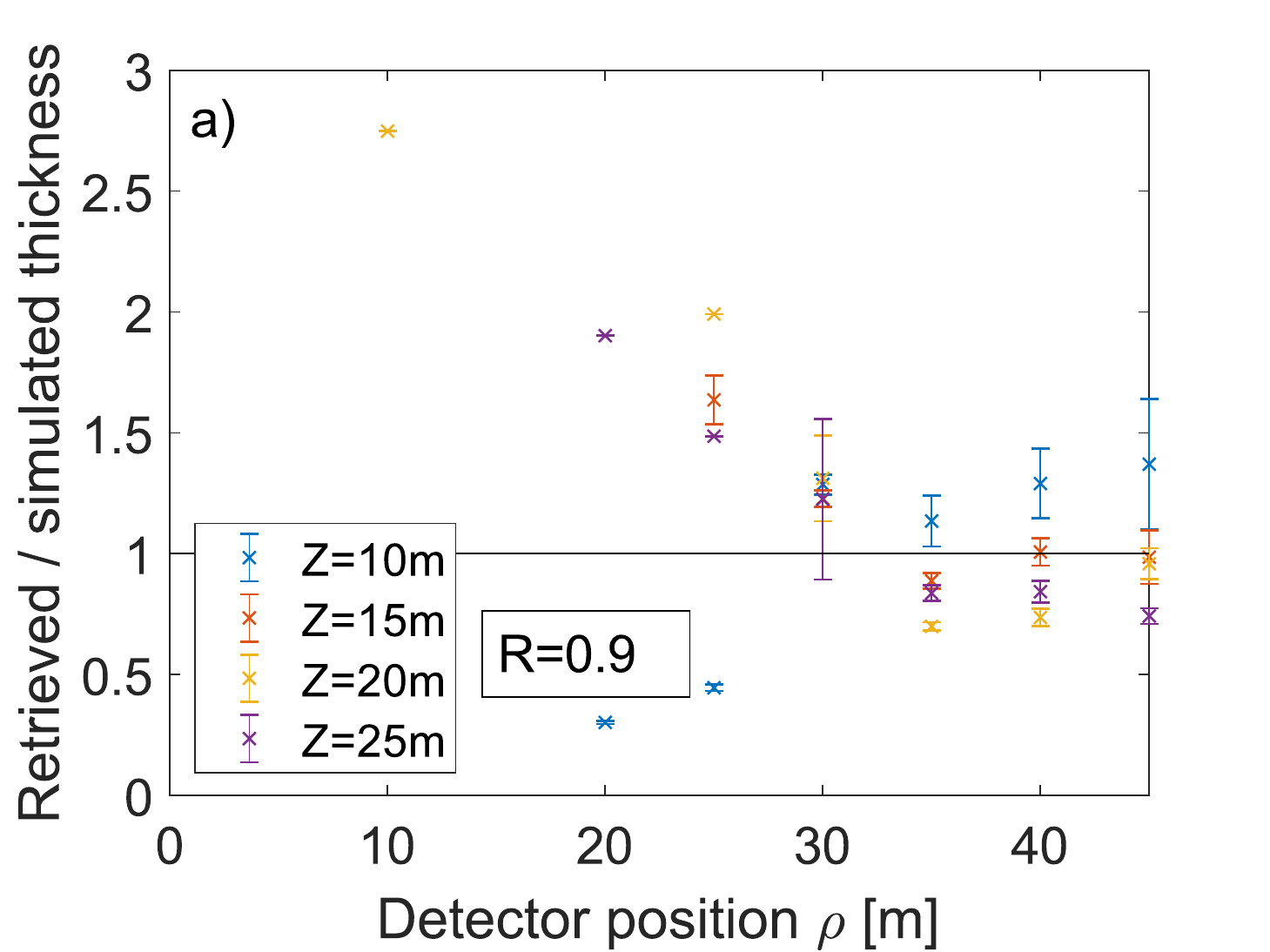}
    \includegraphics[width=0.49\textwidth]{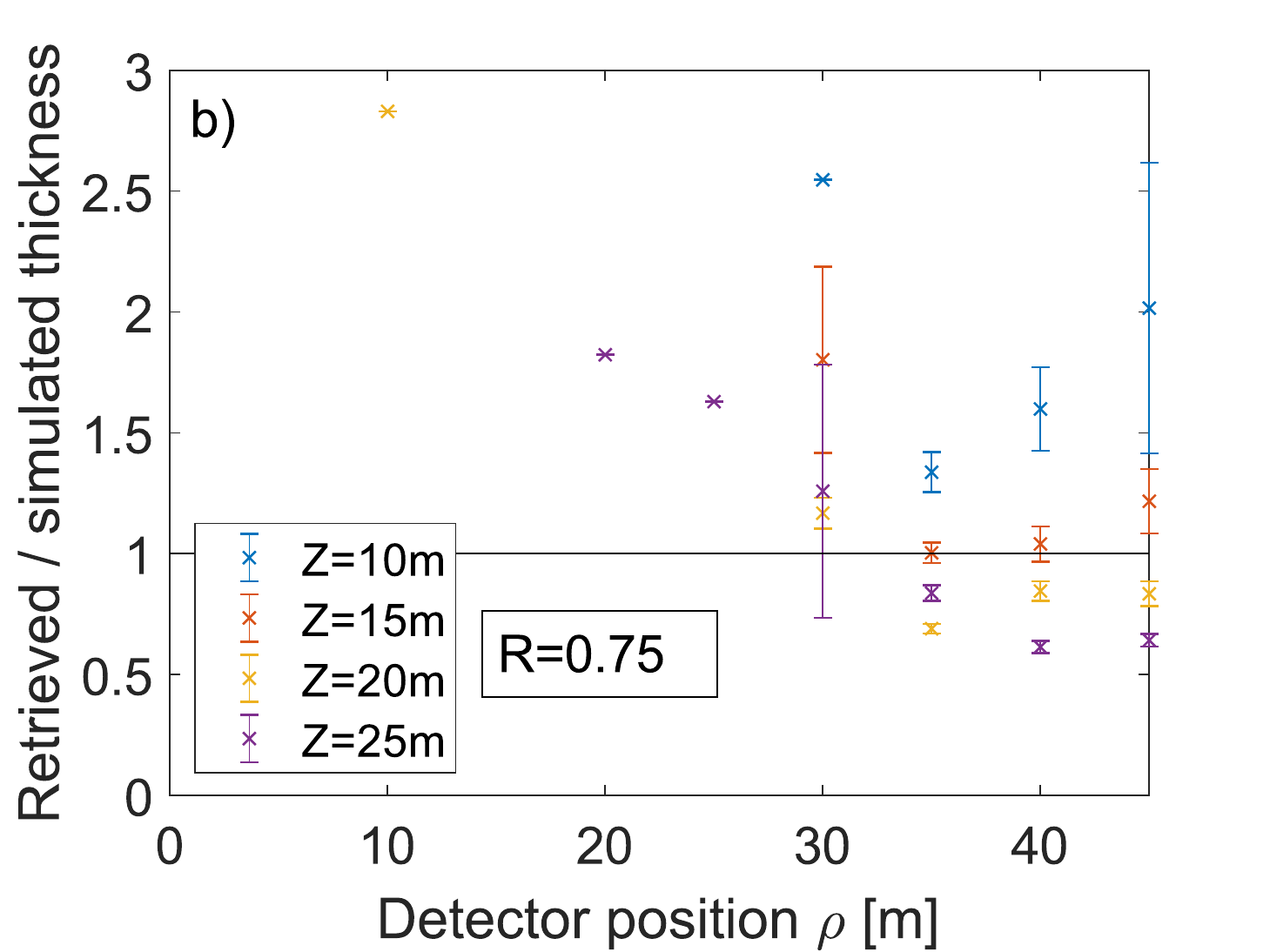}
    \caption{Reconstructed vs simulated volume dept with a lossy lower boundary. Confidence intervals for unreliable fits  are not shown. (a) Lower boundary reflectivity \(R=0.9\).  (b) Lower boundary reflectivity \(R=0.75\).}
    \label{fig:simudepthR}
\end{figure}

Measuring ice thickness is the most challenging of the measurements investigated here, since it is only far away from the source that the volume depth causes a large modification to the diffusion model. Results for volume depths up to 25m are shown in Figure \ref{fig:simudepth}. It can be seen that at small detector separation the thickness is systematically overestimated. However, in those cases the confidence intervals of the fits (not shown where the relative confidence interval is larger than 100\%) are not useful. Retrieved thicknesses trend towards the simulated value at a detector position around 30\,m from the source. At larger separation, fits still converge, but do not converge to the correct value. This can be explained in terms of the theoretical curves shown earlier: At large separation, the signature of the truncated volume turns into what is almost symmetrical around the peak, and is not strongly dependent on the volume thickness anymore. To avoid mismeasurement it seems appropriate to perform the measurement as a function of detector position, and choose the closest separation where fits converge with reasonable confidence. If we assume loss at the lower boundary, parameter retrieval is less reliable, but still possible for moderate loss. We show results for reflectivity \(R=0.9\) and \(R=0.75\) in Figure \ref{fig:simudepthR}. For \(R=0.9\), the fits retrieve thickness values that are comparable to the ones shown previously in Figure \ref{fig:simudepth} for the case of a perfectly reflecting lower boundary. For \(R=0.75\), it can be seen that especially for small source-detector separation, the thickness retrieval fails. As loss is introduced to the lower boundary, the portion of the detectable signal stemming from the reflection is decreased, explaining the loss of this signature. For shallow volumes, this effect is more pronounced because multiple reflections are more likely. In addition, systematic deviations from the simulated volume depth occur, with relative errors of the order of 50\%. We can therefore conclude that the idealized diffusion model can tolerate small amounts of loss of the order of 10\%, but further modeling is needed to account for larger losses at the boundary.

While this approach leaves much to be desired and further investigation is warranted, the results show that thickness retrieval is possible in principle. Askebjer \textit{et al.} \cite{askebjer1997} used a more sophisticated approach for parameter retrieval, where they used curve fitting to only extract maximum and mean of the distribution, and then used the optical parameters extracted from those two figures as a starting point for a maximum likelihood estimation using simulations. However, they did not apply their procedure to retrieval of propagation distances, since those were known. Given that we assumed exact knowledge of optical properties and unit reflectivity of the lower boundary, we conclude that this measurement of ice thickness may be very limited under realistic conditions, and that further research is needed to develop this technique.

To judge their usefulness, these results need to be put into context. The assumptions made about optical properties and access to bare ice definitely restrict the described experiment to the ablation zone of temperate glaciers. Here in Oregon, the thickness of glaciers in the Cascade range was last measured in the 1980's \cite{driedger1986}. At that time, ice thickness rarely exceeded 50m, while some of the glaciers under investigation have since disappeared completely \cite{hartz2020}. The accuracy of depth retrieval demonstrated here is of the orders of 10\%, on par with comparable techniques using radio waves.


\subsection{Experimental considerations}
\label{sec:exp}

Measuring glacier thickness is the most challenging experiment proposed here. The results presented here show that a signature of volume thickness can be extracted in principle and under ideal conditions. While we have also shown that optical properties, which need to be known as precisely as possible to retrieve thickness, can be measured independently, further modeling is needed in order to test the robustness of parameter retrieval. This is particularly true in regard of the lower boundary and realistic structure of the ice. Lastly, if a precise model is needed, the commonly used Henyey-Greenstein  formula as well as mixed phase functions used by others may be insufficient since they do not reproduce the back-scattering peak of the Mie distribution \cite{toublanc96,aartsen2013}.

Bedrock reflection can be as high as 70\,\% for sandstone \cite{watson1971}, 50\,\% for granite and as low as 10\,\% for andesite \cite{zhou2017}. Additional issues such as air and water between the ice and bedrock further complicate formulating an appropriate lower boundary for the diffusion model, since they could either enhance or reduce reflectivity, depending on geometry and optical properties. Therefore, knowledge on local geology and glacier morphology is essential for successful depth measurement. Again, establishing bounds on how robust the diffusion model needs to be for the measurement to be successful will be essential. 

All measurements outlined here will require to measure small quantities of light on photon counting detectors, which will be subject to intrinsic detector dark counts of the order of 10\,\(\mathrm{s}^{-1}\) to 1000\,\(\mathrm{s}^{-1}\). In addition, we have to assume a sizable amount of background light, even at night, which can be somewhat suppressed using spectral filtering around the employed laser wavelength. Both dark and background counts are in principle constant. Using modulation, either by modulating the laser or using a mechanical shutter, changes in those values on scales longer than the modulation period can be accounted for, and detection thresholds of 1\(\mathrm{s}^{-1}\) are possible. In the case of time-resolved measurements, a pulsed laser provides the periodic pulse train. We have shown that detection bin widths of roughly 50\,ns are sufficient to resolve the expected arrival time distributions, which span several microseconds. If a long pulse is used, the measured distribution will be the convolution of the instantaneous source distribution with the laser pulse shape. As long as the pulse duration is much shorter than the detection bin width, the broadening cannot be resolved, and the instantaneous distribution can be used for fitting.

Given the simple nature of these proposed experiments we believe that they can be implemented at relatively low cost. With a photon counting detector, a ns-pulsed diode laser, acquisition electronics and minimal optics for collection, an adequate setup can be implemented with off-the-shelf components for under \$10,000. Stationary measurements are possible at much lower cost. In medical imaging, diffuse optical tomography in the frequency-domain, where a continuous wave laser is modulated at different frequencies, has recently been gaining in popularity \cite{osullivan2012,applegate2020,fantini2020}. This has resulted in further reduction in cost.

\section{Conclusion}

We extended the diffusion model used in AMANDA \cite{askebjer1997} with boundary conditions necessary to describe a semi-infinite volume of glacier ice. The modifications from these boundary conditions make it possible to extract both optical properties and the size of a shallow volume. The stationary fluence encodes information on optical properties.

In the framework of the simplified toy model presented here, the feasibility of several different measurement scenarios using laser sources and photon counting detectors has been demonstrated. Measuring the intensity distribution on the ice surface can offer some insight into scattering and absorption coefficients, but they cannot be measured independently without reference. However, the simplicity of this measurement makes it easy to deploy, potentially enabling a vast increase in data collection that can help refine models for radiative heating of ice and snow. A measurement apparatus can potentially be lightweight and easily fit in a backpack. Especially in the mountain ranges of the western United States, where many glaciers are located in designated Wilderness where motorized access is prohibited, the portability of the measurement apparatus could open new frontiers and enable repeat measurements. While time-of-flight measurements come with the increased effort of coincidence measurements, nanosecond temporal resolution, and high demands on collection efficiency, the measurement apparatus would remain similarly mobile, and can be realized with off-the-shelf components. Not requiring boreholes and being non-invasive would make such an apparatus even easier to deploy in the field. Using diffuse optics and time-of-flight measurements is, to the best of the authors' knowledge, the only viable option to obtain independent measurements of scattering and absorption coefficients, without reference values and assumptions about the nature of scattering processes, and has been proven in the field for deep ice \cite{ackermann2006,aartsen2013}. Moreover, these measurements require only small separation between source and detector of the order of 10 meters, where such measurements appear to insensitive to the volume size. The simple retrieval scheme used here showed systematic overestimation of the scattering length by up to 20\%, and further investigation is needed to remedy this issue. Once optical parameters are known, such measurements are able to reveal the depth of shallow bodies of ice up to 25m in thickness. Since the diffusion description and Monte Carlo simulations are both approximations, further research in the physical nature of the scattering phase function may be needed to facilitate reliable ranging. Future research should also focus both on range of realistic values for bedrock reflectivity and additional modeling including appropriate boundary conditions for absorptive interfaces. While our results imply that diffuse optical ranging cannot compete with radar techniques in range \cite{plewes2001}, an optical measurement device can be built using commercially available components at small cost and size. For the shrinking temperate glaciers in the western United States this technique could serve as an early warning system for dying glaciers and improve hydrological forecasting. Measurement of optical properties of ice in the ablation zone can help to improve radiative transfer models and reveal valuable information on paleoclimate history and local ice properties.\\

\paragraph*{\bf Acknowledgments.}
We thank Anders Carlson and Aaron Hartz of the Oregon Glacier Institute, Stephen G. Warren and Matthew G. Cooper for fruitful discussions that inspired this work.

\paragraph*{\bf Disclosures.}
The authors declare no conflicts of interest.

\paragraph*{\bf Data availability.} Data underlying the results presented in this paper are not publicly available at this time but may be obtained from the authors upon reasonable request.

\bibliography{theory,related,main,properties,boreholes}

\end{document}